\documentclass[12pt]{article}
\usepackage[utf8]{inputenc}
\usepackage[longnamesfirst]{natbib}
\usepackage[colorlinks,citecolor=blue,urlcolor=blue]{hyperref}
\usepackage{epsfig, epsf, graphicx}
\usepackage{color}
\usepackage{float}
\usepackage{caption}
\usepackage{dsfont} % para indicadora
\usepackage{multicol}
\usepackage{subfigure}
\usepackage{amsthm}
\usepackage{amscd}
\usepackage{bbm}
\usepackage{verbatim}   % para hacer comentarios largos
\usepackage{geometry}
\usepackage[Lenny]{fncychap}
\usepackage{pdfpages}
\usepackage{amsmath,amssymb,latexsym}
\usepackage{authblk}

\usepackage{algorithm}
\usepackage{algpseudocode}

\newtheorem{theorem}{Theorem}[section]

\newtheorem{corollary}[theorem]{Corollary}
\newtheorem{proposition}[theorem]{Proposition}

\theoremstyle{definition}

\newtheorem{example}[theorem]{Example}
\newtheorem{assumption}{Assumption}
\theoremstyle{remark}
\newtheorem{remark}{Remark}

{\end{minipage}\end{Sbox}\shadowbox{\TheSbox}}
{\end{minipage}\end{Sbox}\fbox{\TheSbox}}
{\end{minipage}\end{Sbox}\doublebox{\TheSbox}}

\newcommand{\E}{\mathbb{E}}

\newcommand{\Cov}{\mathrm{Cov}}

\begin{document}

\thispagestyle{empty} \baselineskip=28pt \vskip 5mm
\begin{center} {\Huge{\bf Nonparametric Estimation of Functional Dynamic Factor Model }}
	
\end{center}

\baselineskip=12pt \vskip 10mm

\begin{center}\large
%\if1\blind
%{
Israel Mart\'inez-Hern\'andez\footnote[1]{\label{note1} 
\baselineskip=10pt Mathematics and Statistics, 
Lancaster University, Lancaster, 
United Kingdom\\
E-mail: i.martinezhernandez@lancaster.ac.uk}$^{*}$, Jesús Gonzalo\footnote[2]{
\baselineskip=10pt Universidad Carlos III de Madrid, Spain}, 
 and Graciela González-Farías\footnote[3]{
 \baselineskip=10pt Centro de Investigaci\'on en Matem\'aticas, A.C. CIMAT, Mexico \\
 
 \vspace{.1cm}
 This research was partially supported by 1) CONACYT, Mexico, scholarship as visiting research student, 2) CONACYT, Mexico, CB-2015-01-252996 and 3) Centro de Investigaci\'on en Matem\'aticas (CIMAT).}
% } \fi
\end{center}

\baselineskip=17pt \vskip 10mm \centerline{\today} \vskip 15mm

%%%%%%%%%%%%%%%%%%%%%%%%%%%%%%%%%%%%%%%%%%%%%%%%%%%%%%%%%%%%%%%%%%%%%%%%
\begin{center}
{\large{\bf Abstract}}
\end{center}
Data can be assumed to be continuous functions defined on an infinite-dimensional space for many phenomena.  However, the infinite-dimensional data might be driven by a small number of latent variables. Hence, factor models are relevant for functional data. In this paper, we study functional factor models for time-dependent functional data. We propose nonparametric estimators under stationary and nonstationary processes. We obtain estimators that consider the time-dependence property. Specifically, we use the information contained in the covariances at different lags. We show that the proposed estimators are consistent. Through Monte Carlo simulations, we find that our methodology outperforms estimators based on functional principal components. We also apply our methodology to monthly yield curves. In general, the suitable integration of time-dependent information improves the estimation of the latent factors.

\baselineskip=14pt

\par\vfill\noindent
{\bf Some key words:} Functional cointegration; Functional dynamic factor model; Functional time series; $I(1)$ functional process; Long-run covariance operator.
\par\medskip\noindent
{\bf Short title}: Functional Dynamic Factor Model

\clearpage\pagebreak\newpage \pagenumbering{arabic}
\baselineskip=26pt

\section{Introduction}
Functional data analysis (FDA) has attracted interest in recent years in different areas in statistics, where data can be collected almost continuously, e.g., in finance, economics, climatology, medicine, and engineering. FDA is a new methodology and an interesting approach to deal with large-scale, high-dimensional and high-frequency data. Unlike the multivariate approach, which depends on the points at which data are taken, FDA can extract additional information about the continuous underlying stochastic process since each curve is treated as a unit \citep[][]{Ramsay-Silverman2005}. In many real applications, functional data, $\{X_{n}(s); \, s\in D,\, n\in \mathbb{Z}\}$, are time-dependent, e.g., yield curves \citep[][]{Diebold2006}, mortality curves \citep[][]{Hyndmanetal2007},
electricity consumption curves \citep[][]{Dominik}, and intraday price curves \citep[]{KokoszkaH2015}. When the functional data are time-dependent, they are called functional time series \citep[see][for a survey on functional time series]{HorvathK2012}. A way to construct a functional time series is to partition a continuous-time stochastic process $\{\mathcal{Y}_{t},\, t\in \mathbb{R}\}$ into consecutive segments of length $\delta$, that is, $\{X_{n}(s)= \mathcal{Y}_{s}\mathds{1} \{ s\in [n\delta,(n+1)\delta)\},\, n\in \mathbb{Z} \}$, where $\delta$ depends on the dataset application (daily, monthly, annual, etc.). Some functional time series
are naturally modeled by using factor models, e.g., yield curves \citep[][]{Diebold2006}, electricity consumption curves \citep[][]{Dominik}, and intraday price curves \citep[]{KokoszkaH2015}. In this paper, we consider factor models for functional time series. We propose a new methodology to obtain an estimator of the functional dynamic factor (FDF) model.

Factor models represent a large number of dependent variables from a dataset in terms of a small number of latent variables. When the data are time-dependent, attention is focused on dynamic factor models. A dynamic factor model can explain a large fraction of the variance in many macroeconomic series \citep{Giannoneetal2005}, and it is also consistent with broad applications to various phenomena; see \cite{StockWatson88}, \cite{BaiNg2002}, \cite{Bai2003}, \cite{Diebold2006}, \cite{Chloeetal2009}, \cite{Hardleetal2010} and \cite{Keyuretal2011}. Dynamic factor models have been studied for multivariate time series in both stationary and nonstationary cases. \cite{BaiNg2002}, \cite{Bai2003}, \cite{Fornietal2005}, and \cite{LamEtAl2011} considered the stationary case. \cite{StockWatson88} studied factors in a cointegrated time series, \cite{GonzaloGranger95} proposed a method to estimate the factors in a cointegrated time series, \cite{BaiandNg2004} studied the factor structure of large dimensional panels in nonstationarity data, and \cite{PenaPoncela2006} presented a procedure to obtain dynamic factor models for a vector of time series. Here, we also study both cases of functional time series, stationary and nonstationary.

Specifically, we assumed that the functional time series $\{X_{n}\}$ is driven by $K$ latent factor loadings curves $\{F_{1}(s), \ldots, F_{K}(s)\}$ and $K$ latent factor time series $\{\beta_{n,k}\}$, i.e., $X_{n}(s)= \sum_{k=1}^{K} \beta_{n,k} F_{k}(s) + \varepsilon_{n}(s)$. This FDF model was studied  previously by other authors. \cite{HaysSpencerShen2012} assumed that $\{X_{n}(s); s\in D \}$, with $D\subset\mathbb{R}$, is observed in a sample of discrete points, $s\in \{s_{1}, \ldots, s_{m}\}$, and that the factors, $\{\beta_{n,k}\}$, follow an AR$(p)$ model. In that paper, the factor loading $F_{k}(s_{j})$ and the components of the AR$(p)$ model are estimated jointly via maximum likelihood and using the EM algorithm. However, the EM algorithm becomes increasingly complicated when the sample size grows or when the number of point observations of each curve grows. Here, we assume that $X_{n}$ is given in functional form instead of discrete observations of each, and we propose a nonparametric estimator for the latent variables $\{\beta_{n,k}\}$ and $F_{k}$. \cite{Dominik} used the FDF model to forecast electricity spot prices, where the factor loading curves, $F_{k}$, are defined as eigenfunctions of the covariance operator of $X_{n}$, and the factor process $\{\beta_{n,k}\}$ is defined as the corresponding score. 
One disadvantage is that functional principal component analysis (PCA) operates in a static way. That is, if the functional data $\{X_{n}\}$ are time-dependent, then the dynamics are not accurately represented by the principal components, as noted in \cite{HormannEtal2015}. In \cite{JungbackerEtAl2014}, the factor loading curves $F_{k}$ are proposed as cubic splines and rely on hypothesis tests to select the number of knots and their locations. This is inefficient when the sample size increases. \cite{KokoszkaH2015} assumed that the factor loading curves are known and depend on time. 
Then, they propose the use of least square estimators to obtain the factors.

In this paper, we propose a nonparametric estimator of the FDF model, considering time-dependent functional data that are either stationary or nonstationary. The factor processes $\{\beta_{n,k}\}$ are assumed to be scalar processes, and they described the dynamics of the data (by dynamics, we mean the dependence structure over time). The factor loadings  $\{ F_{1}, \ldots,  F_{K}\}$ are assumed to be continuous functions. The subspace generated by the proposed estimators for the factor loading curves represents the dynamics of the functional time series. Thus, our interest is in estimating the trajectory of the factor processes without assuming any model in such a way that each trajectory accurately represents the dependency over time. To take into account the temporal dependency, we consider a specific long-run covariance operator.

The remainder of our paper is organized as follows: In Section \ref{Pre}, we introduce mathematical concepts for functional time series. In Section \ref{FDFM}, we describe the methodology to obtain the estimators in both cases: stationary and nonstationary models. Additionally, we present algorithms and examples to illustrate the methodology. In Section \ref{Properties}, we study the properties and consistency of the proposed estimators. In Section \ref{Simulation}, we evaluate the performance of the proposed estimators under different simulation settings, comparing our results with functional PCA estimators. In Section \ref{Application}, we apply our methodology to the yield curves. Finally, Section \ref{sec:dis} presents a discussion. The proofs are provided in the Appendix.

\section{Preliminaries}\label{Pre}
To describe our methodology, we first introduce some concepts for functional time series. Let $\mathcal{ H}=L^2 (D)$ be the separable real Hilbert space of square integrable functions defined on compact subset $D\subset \mathbb{R}$, with inner product $\langle f, g \rangle = \int_D \!f(x)g(x)\, \mathrm{d}x $, and the corresponding norm denoted by $\|\cdot \|_{\mathcal{H}}$. A functional random variable is defined as a random variable in $ \mathcal{ H}$, i.e., $X:(\Omega, \mathcal{F}, \mathbb{P}) \to \mathcal{ H}$.
Let $L_{\mathcal{ H}}^p= \left\{ X: \Omega \to \mathcal{ H} \, : \, \E (\| X \|^p) < \infty   \right\}$ be the set of random variables in $\mathcal{ H}$ with finite moments of order $p$. The expected value of $X\in L_{\mathcal{ H}}^1$ is defined as a unique element of $\mathcal{ H}$, denoted by $\mu$, such that $\E \langle X , y \rangle =  \langle \mu , y \rangle,  $
for all $y \in \mathcal{ H}$. In the rest of the paper, we write $\E(X)$ instead of $\mu$ to refer to the expected value of $X$.

A functional time series is a sequence of functional random variables $\{X_{n},\, -\infty < n < \infty\}$ in $\mathcal{ H}$. The covariance operator at lag $h$ of $\{X_{n}\}\in L_{\mathcal{ H}}^2$ is defined by $\Gamma_{X_{n}, X_{n+h}}(z)= \E \left[ \langle X_{n} - \E(X_{n}), z \rangle \{ X_{n+h}- \E(X_{n+h}) \}\right]$ for all $z\in \mathcal{H}$. This covariance operator can be written as
$\Gamma_{X_{n}, X_{n+h}}(z)(s)= \int\! \gamma_{n, n+h}(t,s )z(t) \mathrm{d}t,$ where $ \gamma_{n, n+h}(t,s)=\mathrm{Cov}(X_{n}(t), X_{n+h}(s))$ is called the kernel of $\Gamma_{X_{n}, X_{n+h}}$. A functional time series $\{X_{n}\} \in L_{\mathcal{ H}}^2 $ is said to be stationary if (i) $\E(X_{n})= \E(X_{0})$ and (ii) $\Gamma_{X_{n+h}, X_{m+h}}= \Gamma_{X_{n}, X_{m}}$ for all $n$, $m$, and $h$. In this case, we use the notation $\Gamma_{h}$ instead of $\Gamma_{X_{n}, X_{n+h}}$. If $\{X_{n}\}$ is a stationary functional time series with $\E(X_{n})=0$ and $\Gamma_{h}=0$ for all $h\neq 0$, then it is called \textit{functional white noise} and \textit{strong functional white noise} if it is a sequence of i.i.d. functional random variables. In the rest of the paper, we refer to strong functional white noise as an i.i.d. sequence in $L_{\mathcal{ H}}^2$. The reader can consult \cite{Bosq2000} and \cite{HorvathK2012} for more details on functional time series.

Let $\{X_{n},\, n\in \mathbb{Z}\}$ be a stationary functional time series, and the long-run covariance operator $\Gamma$ of $X_{n}$ is defined as
\begin{equation}\label{LongRunCov}
\Gamma(z) (s)= \int_{D} \! c(t,s)z(t) \mathrm{d}t, \,\, z\in \mathcal{H}
\end{equation}
where the corresponding kernel is $c(t,s) =\sum_{j=-\infty}^{\infty} \gamma_{j}(t,s)$, and $ \gamma_{j}(t,s)=\mathrm{Cov}(X_{0}(t), X_{j}(s))$. We should note that the assumption of stationarity on $\{X_{n}\}$ does not guarantee the existence of $\Gamma$. For that, we need an additional weak dependence condition stated in Section \ref{Properties}.

Let $\mathcal{B}_{\mathcal{H}}= \{A: \mathcal{H} \to \mathcal{H}; A \mbox{ is continuous and linear} \}$ be the set of all continuous linear operators from $\mathcal{H}$ to $\mathcal{H}$, with an operator norm denoted by $\|\cdot \|_{\mathcal{B}_{\mathcal{H}}}$. 
Let $A\in \mathcal{B}_{\mathcal{H}}$; an eigenfunction $v$ of $A$ is defined as a nonzero element of $\mathcal{H}$ such that $A(v )= \alpha v,$ where $\alpha\neq 0$ is a scalar number and is the associated eigenvalue.

Let $\{ \varepsilon_{n} \}$ be an i.i.d. sequence in $L^{2}_{\mathcal{H}}$, and let $\{A_{n}\} \in \mathcal{B}_{\mathcal{H}}$. A functional linear process $\{Y_{n}, n\in \mathbb{Z}\}$ with innovations $\{ \varepsilon_{n} \}$ is defined as
\begin{equation*}\label{LinOp}
Y_{n}(s)=\sum_{j=0}^{\infty} A_{j}(\varepsilon_{n-j})(s), \quad s\in D.
\end{equation*}
If the functional linear process $\{Y_{n}\}$ is stationary, and if its covariance long-run covariance operator exists, then this  long-run covariance operator can be written as (see Appendix)
\begin{equation}\label{LongRCovLP}
\Gamma = A\Gamma_{\varepsilon_{0}} A^{*},
\end{equation}
where $\Gamma_{\varepsilon_{0}} $ is the covariance operator of $\varepsilon_{0}$, $A=\sum_{j=0}^{\infty}A_{j}$ and $ A^{*}$ is the adjoint operator of $A$.

Now, we introduce the functional $I(1)$ process. We use the approach proposed by \cite{BeareEtal2017}, where the notion of cointegration for multivariate time series is extended to an inifinite-dimensional space. Let $\{X_{n}\}\in L_{\mathcal{ H}}^2$ be a functional time series such that the first difference $\Delta X_{n}=X_{n}- X_{n-1}$ admits the representation $\Delta X_{n}=\sum_{j=0}^{\infty} \Phi_{j}(\varepsilon_{n-j})$, where $\{\varepsilon_{n}\}$ is an i.i.d. sequence in $L_{\mathcal{ H}}^2$, and coefficient operators satisfying $\sum_{j=0}^{\infty}j \| \Phi_{j}\|_{\mathcal{B}_{\mathcal{H}}} < \infty$. Then, $X_{n}$ can be written as $X_{n}= X_{0}+ \Phi \left(\sum_{j=1}^{n} \varepsilon_{j}\right) + \eta_{n}$, where $\Phi = \sum_{j=0}^{\infty} \Phi_{j}$, and $\{\eta_{n}\}$ is a stationary functional time series. The functional time series $\{X_{n}\}$ is called an $I(1)$ functional process if and only if the long-run covariance operator of $\{\Delta X_{n}\}$ is different from zero.

\section{Methodology}\label{FDFM}
In this paper, we assume that the functional time series $\{ X_{n} \}$ are given in the functional form. In a real application, functional data are observed on a grid of points, and thus, the continuous curve should be estimated \citep[see][Chapter 3-7]{Ramsay-Silverman2005}.

\subsection{\textit{ Model setting}} \label{MS}

Assume that we observe $N$ functional data $\{X_{1}, \ldots, X_{N} \}$. We assume that the functional data follow the model
\begin{equation}\label{Funct-Model}
 \begin{split}
&X_{n}(s) =  Y_{n}(s)+ \varepsilon_{n}(s),\\
&Y_{n}(s) =\sum_{k=1}^{K}  \beta_{n,k} F_{k} (s)
 \end{split}
\end{equation}
where $F_{k}$, $k=1, \ldots, K$, are factor loading curves, $\{ \beta_{n,k}\}$ are scalar factor time series, $K$ is the number of factors, and $\{\varepsilon_{n} \}$ is a sequence of centered, independent and identically distributed innovations in $L^2_{\mathcal{H}}$ with covariance operator $\Gamma_{\varepsilon}$. We refer to model $\eqref{Funct-Model}$ as the functional dynamic factor model or FDF model. The factors $\{\beta_{n,k}\}$ are assumed to be time-dependent, and then $\{X_{n}\}$ is time-dependent. The $K$ factor processes drive the dynamics of the functional data $\{X_{n}\}$. Factors and factor loadings are latent functions and latent random variables, respectively.

For model identifiability, we assume orthonormality of the factor loading curves, that is,
\begin{equation}\label{ortho}
\int_{D}\! F_{i}(s)  F_{j}(s)\mathrm{d}s = \mathds{1}( j=i), 
\end{equation}
for $i, j = 1, \ldots, K$, where $\mathds{1}(j=i)$ takes the value $1$ if $i=j$ and zero otherwise. In some scenarios, the factor processes $\{\beta_{n,k} \}$ can be known, and then, condition \eqref{ortho} can be omitted \citep[][]{KokoszkaEtal2018}.

Since factors and factor loadings are unobserved, $F_{k}$ and $\{ \beta_{n,k}\}$ are not uniquely determined in model \eqref{Funct-Model}, even with the constraint \eqref{ortho}. However, the linear space $\mathcal{H}_{F}:= \mathrm{span}\{F_{1}, \ldots, F_{K}\}$ generated by the factor loadings, called the factor loading space, is uniquely defined. Thus, any orthonormal rotation of the orthonormal basis system $\{F_{1}, \ldots, F_{K} \}$ can be a solution to model \eqref{Funct-Model} as well.  Therefore, our goal is to estimate $\mathcal{H}_{F}$. To do so, we assume the following conditions.

\begin{assumption}\label{A1} The functional white noise $\{\varepsilon_{n}\}$ is uncorrelated with the functional process $\{ Y_{n}\}$, that is, $\Gamma_{Y_{n}, \varepsilon_{n+h} }=0$, $h \in \mathbb{Z}$.
\end{assumption}
\begin{assumption}\label{A3} There exists $i,j\in \{1,\ldots, K\}$ such that $\mathrm{Cov} (\beta_{n,i}, \beta_{n, + h, j}) \neq 0$, for some $h>0$.
\end{assumption}

Assumptions \ref{A1} is the usual conditions assumed for the FDF model. Assumption \ref{A3} requires time-dependent functional data $\{X_{n}\}$ and ensures $\widehat{\mathcal{H}}_{F}$ not being an empty set. To motivate our methodology, let us assume that the functional data $\{X_{n}\}$ are stationary and follow model \ref{Funct-Model} (we will use similar ideas for the nonstationary case in Section \ref{NScase}). Thus, under Assumption \ref{A1}, the covariance operator of $X_{n}$ satisfies 
$$\Gamma_{h}= \sum_{i,j }^K  \Cov(\beta_{n,i}, \beta_{n+h,j})   F_{i}\otimes  F_{j} +  \mathds{1}(h=0) \Gamma_{\varepsilon}, \quad h\in \mathbb{Z},$$
where $\otimes$ denotes the tensor product. Therefore, $\Gamma_{h} (v) =0$ for any $v\in\mathcal{H}_{F}^{\bot}$, with $h\neq 0$. Moreover, if $\mathrm{ker} ( \Gamma_{0})= \{z\in \mathcal{H}: \Gamma_{0} (z)=0 \}= \{0\}$ and $\Gamma_{0}^{-1}$ is invertible, then $\mathrm{ker}( \Gamma_{h} \Gamma_{0}^{-1})= \{\mathrm{range}( \Gamma_{h}^{*}) \}^{\bot} = \mathcal{H}_{F}^{\bot}$. Hence, $\mathcal{H}_{F}$ is the orthogonal complement of the linear space spanned by the eigenfunctions of $\Gamma_{h} \Gamma_{0}^{-1}$ (or $\Gamma_{h}$) corresponding to the zero eigenvalues, with $h\neq 0$. Thus, our methodology uses the operator $\Gamma_{h} \Gamma_{0}^{-1}$, and we consider the summation over all possible lags $h\neq 0$, that is,
\begin{equation}\label{Op_Lambda}
\Lambda := \sum_{h\in \mathbb{Z},h\neq 0} \Gamma_{h} \Gamma_{0}^{-1}= (\Gamma - \Gamma_{0})\Gamma_{0}^{-1},
\end{equation}
where $\Gamma$ is the long-run covariance of $X_{n}$ defined in \eqref{LongRunCov}. We notice that, under Assumption \ref{A3},  $\mathrm{range} (\Lambda) \neq \emptyset$.

We observe that $\Lambda(v)=0$ for any $v\in  \mathcal{H}_{F}^{\bot}$ as well. Consequently, we propose to estimate the space $ \mathcal{H}_{F}$ using the eigenfunctions of the operator $\Lambda$ corresponding to nonzero eigenvalues.  Additionally,  these eigenfunctions are defined as estimators of the factor loadings $F_{k}$, $k=1, \ldots, K$.

In general, $\Gamma_{0}^{-1}$ is unbounded \citep[][]{Bosq2000, MARTINEZHERNANDEZ2019}. The method typically employed to address this problem is to truncate the spectral representation of $\Gamma_{0}$ and then compute the inverse from this representation. We adopt this method, and we describe it in Section \ref{StationaryCase}.

\begin{remark} The consideration of the term $\Gamma_{0}^{-1}$ in \eqref{Op_Lambda} is equivalent to assuming that the data have identity as the covariance operator.
  The latter case requires transforming the data $X_{n}$ to $\Gamma_{0}^{-1/2} (X_{n})$. Thus, $\Gamma_{h} \Gamma_{0}^{-1}$ is the covariance operator at lag $h$ of
the transformed data.
\end{remark}

\subsection{\textit{Estimation for stationary FDF model}}\label{StationaryCase}
Without loss of generality, we assume that $\mathbb{E} (X_{n})=0$. To estimate $\mathcal{H}_{F}$, we compute the eigenfunctions of $\widehat{\Lambda}$, where $\widehat{\Lambda}$ is the estimator of $\Lambda$. To obtain $\widehat{\Lambda}$, we use a smooth periodogram-type estimator \citep{GregoryAndShang2017}. Let $\widehat{\gamma}_{h}(t,s)$ be the kernel estimator of $\Gamma_{h}$ defined as
\begin{equation*}
 \widehat{\gamma}_{h}(t,s)= \left\{ \begin{array}{lcl} 
 \frac{1}{N}\sum_{i=1}^{N-h}  X_{i}(t)  X_{i+h}(s), &  h\geq 0  \\
 \frac{1}{N}\sum_{i=1-h}^{N}  X_{i}(t) X_{i+h}(s), &  h< 0  
  \end{array}
 \right. 
 \end{equation*}
Now, we define $\widehat{c}(t,s)- \widehat{\gamma}_{0}(t,s)$ as
$$\widehat{c}_{b,-0}(t,s)=  \sum_{|h| \leq b,\, h\neq 0}\chi \left (\frac{h}{b}\right ) \widehat{\gamma}_{h}(t,s) ,$$
where $\chi$ is a continuous, symmetric weight function and satisfies $\chi(0)=1$, $\chi(u)=0$ if $| u |>c$ for some $c>0$. In this paper, we use $\chi (h/b ) = 1- |h|/b$, where $b$ is a bandwidth parameter. To obtain a consistent estimator, the bandwidth must satisfy $b=b(N)\to \infty $ as $N\to \infty$ and $b(N)/ N= o(1)$ \citep[][]{HorvathEtAl2012}. However, in practice, the selection of $b$ should be done carefully, since this can affect the performance of the estimator in finite samples. Here, we select $b$ similarly as described in \cite{GregoryAndShang2017}, that is,
minimizing the mean-squared error $\mathbb{E} \| \widehat{c}_{b} - c \|^{2}$, where $\| \cdot \|$ denotes the usual norm in $L^{2}([0,1]^{2})$. We use the notation $\widehat{c}_{-0}(t,s)$ to denote $\widehat{c}_{b,-0}(t,s)$ with the optimal value of the bandwidth $b$.

Then, the estimator of the operator $\Gamma - \Gamma_{0}$ is defined by $\widehat{\Gamma - \Gamma_{0}}(z)(s)=\int \! \widehat{c}_{-0}(t,s) z(t)\, \mathrm{d}t $, $z\in \mathcal{H}$. To obtain a bounded estimator of $\Gamma_{0}^{-1}$, we only consider the first $p$ eigenfunctions corresponding to the largest eigenvalues of $\Gamma_{0}$. Thus, the inverse $\widehat{\Gamma}_{0}^{-1}$ is approximated by $\widehat{\Gamma}_{0}^{-1} = \sum_{j=1}^{p} \widehat{\lambda}_{j}^{-1} \widehat{v}_{j} \otimes \widehat{v}_{j}$, where $ \widehat{v}_{j}$, $j=1, \ldots, p$, are such that $\widehat{\Gamma}_{0} (v_{j})= \widehat{\lambda}_{j} \widehat{v}_{j}$. We select the parameter $p= p(N)$ using the scree plot.

Finally, we obtain an estimator of $\Lambda$ as $\widehat{\Lambda} = \widehat{\Gamma - \Gamma_{0}}  \widehat{\Gamma}_{0}^{-1} $, and consecutively, we estimate the eigenfunctions $\{\widehat{\zeta}_{1}, \ldots, \widehat{\zeta}_{k_{0}}\}$ of $\widehat{\Lambda}$ corresponding to  the first $k_{0}$ largest nonzero eigenvalues $\{\widehat{\alpha}_{1}, \widehat{\alpha}_{2}, \ldots,  \widehat{\alpha}_{k_{0}}\}$, with $k_{0}$ a positive number. That is, for each $i=1, \ldots, k_{0}$, $\widehat{\zeta}_{i}$ satisfies
$$\widehat{\Lambda} \widehat{\zeta}_{i}= \widehat{\alpha}_{i} \widehat{\zeta}_{i}.$$

As mentioned above, we have that  $K$ is the number of nonzero eigenvalues of $\Lambda$. In practice, the number of eigenfunctions of $\widehat{\Lambda}$ with nonzero eigenvalues is not exactly $K$, since the zero-eigenvalues of $\widehat{\Lambda}$ are unlikely to be zero exactly. Here, we estimate the number of factors similarly as in \cite{LamEtAl2012}, using a ratio-based estimator. Explicitly, we define $\widehat{K}$ as
\begin{equation}\label{EstNumF}
\widehat{K}= \arg \min_{1\leq i\leq k_{0}} \widehat{\alpha}_{i+1} / \widehat{\alpha}_{i},
\end{equation}
where $k_{0}$ should be large enough that $K<k_{0}$.  Although we do not study the theoretical properties of $\widehat{K}$, thorough empirical investigations suggest it works well in practice.

Once we have estimated the eigenfunctions of $\Lambda$ and the number of factors, we define $\widehat{\mathcal{H}}_{F}= \mathrm{span} \{\widehat{\zeta}_{1}, \ldots, \widehat{\zeta}_{\widehat{K}}\}$, and $\widehat{F}_{i} = \widehat{\zeta}_{i}$, $i=1,\ldots, \widehat{K}$. The estimated trajectories of the factor processes are obtained as
$$\{\widehat{\beta}_{n,k}\}_{n=1}^{N} = \{\langle X_{n}, \widehat{F}_{k} \rangle \}_{n=1}^{N}, \quad k=1, \ldots, \widehat{K}.$$

 Algorithm \ref{A1} presents a summary of the steps to obtain the estimators of model \eqref{Funct-Model} under the stationary assumption.
\begin{algorithm}
\caption{Estimators for stationary FDF model} \label{A1}
\label{array-sum}
\begin{algorithmic}[1]
\State Fix $k_{0}$ with $k_{0}$ large enough.
\State Compute the estimator $\widehat{\Lambda}$.
\State Compute the eigenfunctions $\widehat{\zeta}_{1}, \ldots, \widehat{\zeta}_{k_{0}} $ of $\widehat{\Lambda}$, corresponding to the first $k_{0}$ largest eigenvalues.
\State Obtain $\widehat{K}$ as in \eqref{EstNumF}.
\State For $k=1,\ldots, \widehat{K}$, estimate the factor loadings and the trajectories of the factors as follows: $\widehat{F}_{k}= \widehat{\zeta}_{k}$, and $\{\widehat{\beta}_{n,k}, \, n\geq 1\}= \{\langle X_{n},\widehat{\zeta}_{k} \rangle, \, n\geq 1 \} $.
\end{algorithmic}
\end{algorithm}

\begin{figure}[!b]
\begin{center}
\includegraphics[scale=.41]{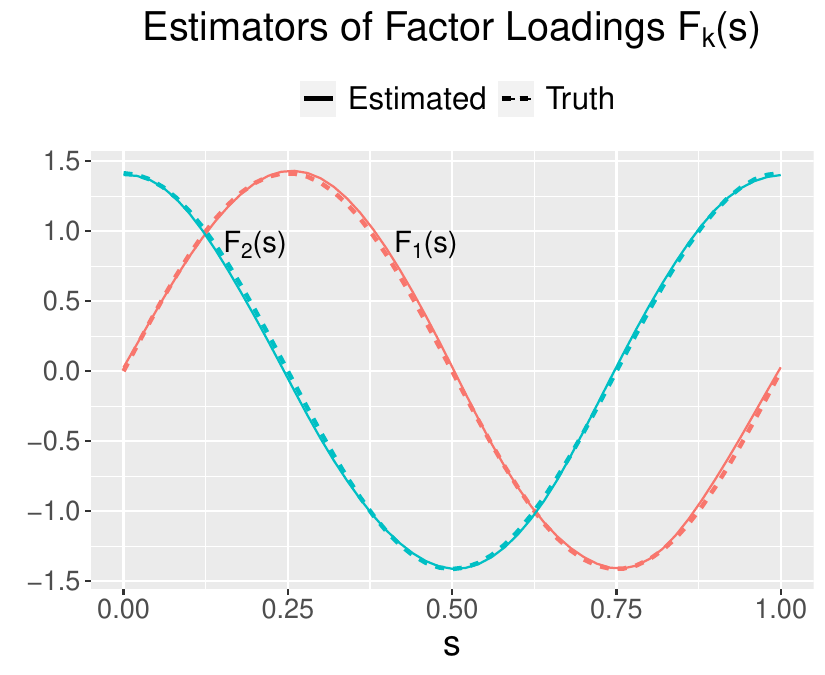}\hspace{-.2cm}
\includegraphics[scale=.41]{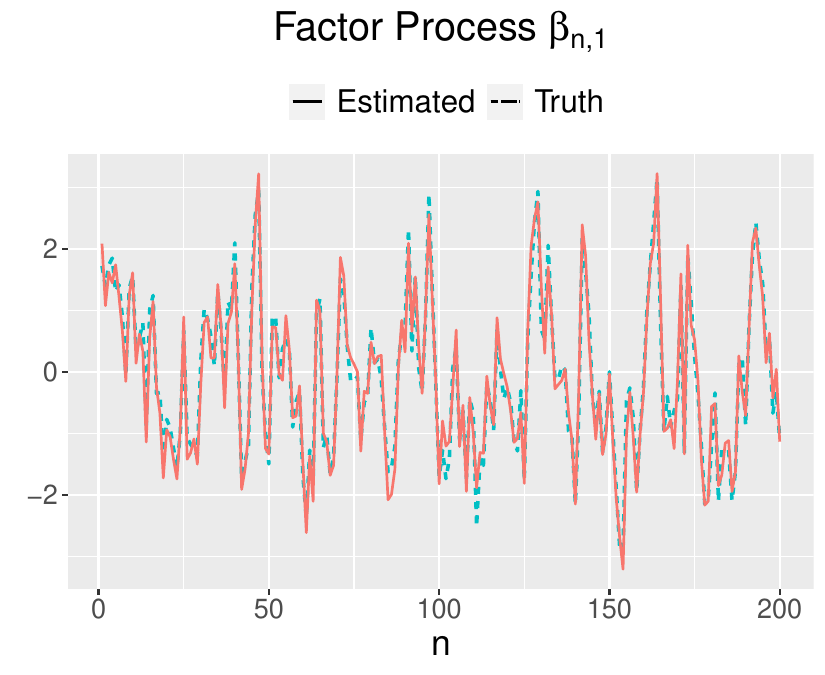}\hspace{-.2cm}
\includegraphics[scale=.41]{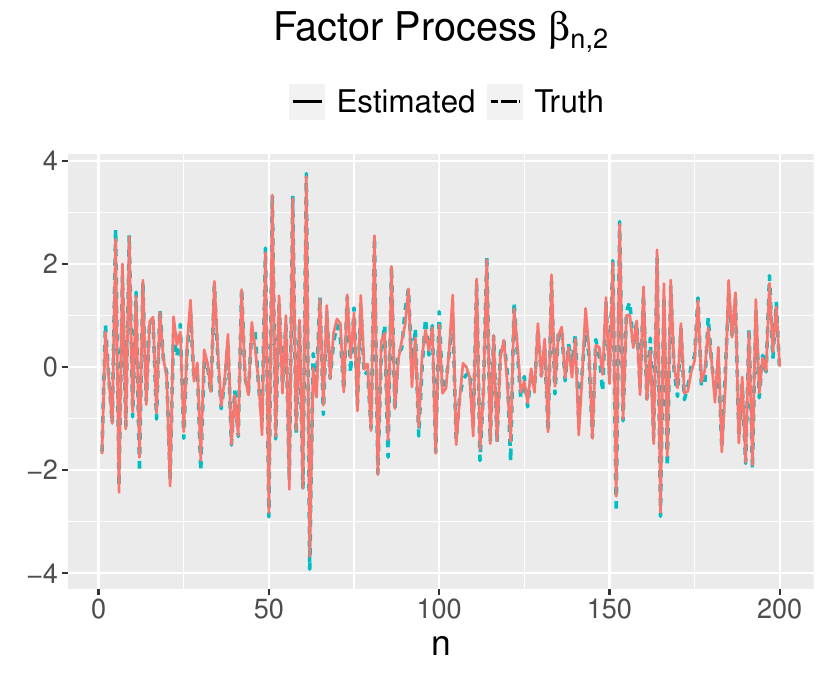}
%\vspace{-.5cm}
\caption{Estimators of $F_{1}(s)$ and $F_{2}(s)$ for the stationary FDF model with $K=2$ and the estimated trajectories for the factor time series $\{\beta_{n,1}\}$ and $\{\beta_{n,2}\}$. The sample size is $N=200$.} \label{ExampleEstimators}
\end{center}
\end{figure}
\begin{example}\label{Example1} We simulate $\{X_{n}\}$ from an $\mathrm{FDF}$ model with $K=2$ and sample size $N=200$. The factors are AR$(1)$ processes with coefficients $a_{1,1}=0.6$ and $a_{2,1}=-0.6$. The factor loadings are defined by $F_{1}(s)=\sin(2\pi s)$ and $F_{2}(s)=\cos(2\pi s)$ with $s\in [0,1]$. Figure \ref{ExampleEstimators} shows the estimators obtained from Algorithm \ref{A1}. The left panel shows the factor loadings and their corresponding estimators, while the center and right panels show the estimated trajectory for the factors. We observe that $\widehat{F}_{i}$ is close to the original factor loadings and that the estimated trajectories of the factor processes successfully approximately the original factor processes. Thus, the proposed estimators have good performance in this example. A more exhaustive simulation study is performed in Section \ref{Simulation}.
\end{example}

\subsection{\textit{Estimation for the nonstationary FDF model}}\label{NScase}

In real data applications, the factor processes can be nonstationary. Here, we assume that at least one of the factor processes $\{\beta_{n,k} \}$, $k=1, \ldots, K$, is an $I(1)$ scalar process. Let $r\in \{1, \ldots, K\} $ be the number of factor processes that are $I(1)$ processes. We consider two cases: (i) all factor processes are $I(1)$ processes, i.e., $r=K$, and (ii) there are $r$ $I(1)$ factor processes, with $r<K$, and the remaining $K-r$ factor processes are stationary processes. 
Thus, the factor loading space can be written as $\mathcal{H}_{F} = \mathcal{H}_{F}^S \oplus \mathcal{H}_{F}^N$, where $ \mathcal{H}_{F}^S $ is the linear space generated by the stationary factor processes and $ \mathcal{H}_{F}^N$ is the linear space generated by the nonstationary factor processes. We have that if $z\in  \mathcal{H}_{F}^S $, then the scalar time series $\{ \langle X_{n}, z \rangle \}_{n}$ is stationary, and if $z\in  \mathcal{H}_{F}^N $, the scalar time series $\{ \langle X_{n}, z \rangle\}_{n}$ is nonstationary. Case (ii) is related to the concept of cointegrating in a Hilbert space \citep{BeareEtal2017}. Here, we describe estimators for $ \mathcal{H}_{F}^S $ and $ \mathcal{H}_{F}^N $.

Let $\Lambda_{\Delta X}$ be the operator defined in \eqref{Op_Lambda} for the functional process $\Delta X_{n}$. In the following, we describe the two cases for the factor processes.

\textit{Case (i)}: In this case, $ \mathcal{H}_{F} =  \mathcal{H}_{F}^N$, and $\Delta X_{n}$ is a stationary FDF model. We estimate the space $ \mathcal{H}_{F}^N$ as the linear space generated by the eigenfunctions of the operator $\Lambda_{\Delta X}$, where $\Lambda_{\Delta X}$ is estimated as described in Section \ref{StationaryCase} using $\{ \Delta X_{n}, n\geq 1\}$ instead of $\{X_{n}, n\geq 1\}$. The number of factors is estimated as in \eqref{EstNumF} with the corresponding eigenvalues of $\widehat{\Lambda}_{\Delta X}$. Finally, we define $\widehat{F}_{k}=\widehat{\xi}_{k}$, $k=1, \ldots, \widehat{K}$,
where $\widehat{\xi_{k}}$ are the eigenfunctions of $\widehat{\Lambda}_{\Delta X}$. This approach guarantees that
$\{\widehat{\beta}_{n,k},\, n\geq 0\}= \{ \langle X_{n}, \widehat{\xi}_{k} \rangle, \, n\geq 0 \}$ is an $I(1)$ process for $k=1, \ldots, \widehat{K}$ (Proposition \ref{Prop:CS}).

\textit{Case (ii)}: We have that $\mathcal{H}_{F} = \mathcal{H}_{F}^S \oplus \mathcal{H}_{F}^N $ with $ \mathcal{H}_{F}^{S} \neq \emptyset$. First, we estimate the space $\mathcal{H}_{F}^N$; then, we subtract the estimated space $\widehat{\mathcal{H}}_{F}^{N}$ from the entire space $\mathcal{H}$. Then, we estimate $\mathcal{H}_{F}^S$. We propose to estimate $\mathcal{H}_{F}^N$ as the linear space generated by the eigenfunctions of $\Lambda_{\Delta X}$, and then we define $\widehat{F}_{k}=\widehat{\xi}_{k}$, where $\widehat{\xi}_{k}$ is the eigenfunction of $ \widehat{\Lambda}_{\Delta X}$, for $k=1, \ldots, \widehat{r}$. Here, we obtain $\widehat{r}$ using criteria  \eqref{EstNumF}, with $\widehat{\alpha}_{i}'s$ being eigenvalues of $ \widehat{\Lambda}_{\Delta X}$. Then, we obtain the estimated trajectories of the factor processes as $\{\widehat{\beta}_{n,k},\, n\geq 0\}= \{ \langle X_{n}, \widehat{\xi}_{k} \rangle, \, n\geq 0 \}$, for $k=1, \ldots, \widehat{r}$. Given $\{\widehat{F}_{1}, \dots, \widehat{F}_{\widehat{r}} \}$, we estimate $ \mathcal{H}_{F}^S$ as follows. We define a new functional time series as
\begin{equation}\label{FDFS}
Z_{n}(s)= X_{n}(s)- \sum_{k=1}^{\widehat{r}} \widehat{\beta}_{n,k} \widehat{F}_{k}(s).
\end{equation}
Let $\Lambda_{Z}$ be the corresponding operator defined in \eqref{Op_Lambda} for the functional process
$\{Z_{n}\}$. The functional process $\{Z_{n}\}$ should be a stationary functional process. Thus, we estimate $\mathcal{H}_{F}^S$ similarly as in Section \ref{StationaryCase}, with $\widehat{K-r}$ as in \eqref{EstNumF} using the corresponding eigenvalues of $\widehat{\Lambda}_{Z}$. Let $\{ \widehat{\upsilon}_{1}, \ldots, \widehat{\upsilon}_{\widehat{K-r}} \}$ be the eigenfunctions of $\widehat{\Lambda}_{Z}$ corresponding to the $i$-th largest eigenvalues. Then, we define $\widehat{F}_{\widehat{r}+1}= \widehat{\upsilon}_{1}, \ldots,\widehat{F}_{\widehat{K}}= \widehat{\upsilon}_{\widehat{K-r}} $, with $\widehat{K}= \widehat{r} + \widehat{K-r}$, as the estimators of the factor loadings corresponding to the stationary dynamic and $\{\widehat{\beta}_{n,k},\, n\geq 0\}= \{ \langle X_{n}, \widehat{\upsilon}_{k-r} \rangle, \, n\geq 0 \}$, for $k=r+1, \ldots, K$, as the estimated trajectories of the $I(0)$ factor processes.

In real applications, we do not know whether $r<K$ or $r=K$. To overcome this problem, we propose to apply a test for independence for the functional process $\{Z_{n}, n\geq 1\}$, such as a test proposed in \cite{HorvathEtAl2013}. If the independency hypothesis is not rejected, then we conclude that $r=K$. Otherwise,  $r<K$, and we can proceed to estimate the corresponding stationary loading space $\mathcal{H}_{F}^S$.

In Algorithm \ref{A2}, we present a summary of the steps to obtain the estimators of model \eqref{Funct-Model} with $r$ nonstationary factor processes and $K-r$ stationary factor processes.

\begin{algorithm}
\caption{Estimators for nonstationary FDF model} \label{A2}
\label{array-sum}
\begin{algorithmic}[1]
\State Apply Algorithm \ref{A1} by considering $\Lambda_{\Delta X}$, and define $\widehat{r}= \widehat{K}_{\Delta X}$, where $ \widehat{K}_{\Delta X}$ is the number of factors estimated with eigenvalues of $\widehat{\Lambda}_{\Delta X}$.
\State Save the estimated loading factors and factor processes, $\{\widehat{F}_{k}\}$ and $\{ \widehat{\beta}_{n,k}\}$, $k=1, \ldots, \widehat{r}$.
\State Obtain the functional process $\{Z_{n}\}$ as defined in \eqref{FDFS}.
\State Apply a test of independence for the functional process $\{Z_{n}\}$.
\State  If the test of the independency hypothesis is not rejected, then stop.
\State Else: Compute $\widehat{\Lambda}_{ Z}$ the estimator of the operator $\Lambda_{Z}$ corresponding to the functional process \eqref{FDFS}.
\State Apply Algorithm \ref{A1} to $\widehat{\Lambda}_{Z}$, and define $\widehat{K-r}=\widehat{K}_{Z}$, where $\widehat{K}_{Z}$ is the number of factors estimated with eigenvalues of $\widehat{\Lambda}_{Z}$.
\State Finally, obtain the complete estimators of loading factors and factor processes, $\{\widehat{F}_{1}, \ldots, \widehat{F}_{\widehat{r}}, \widehat{F}_{\widehat{r}+1}, \ldots, \widehat{F}_{\widehat{K}}\}$ and $\{ \widehat{\beta}_{n,1} , \ldots, \widehat{\beta}_{n,\widehat{r}}, \widehat{\beta}_{n,\widehat{r}+1}, \ldots, \widehat{\beta}_{n,\widehat{K}} \}$, where $\widehat{K} = \widehat{r} + \widehat{K-r}$.
\end{algorithmic}
\end{algorithm}

\begin{figure}[!b]
\begin{center}
\includegraphics[scale=.41]{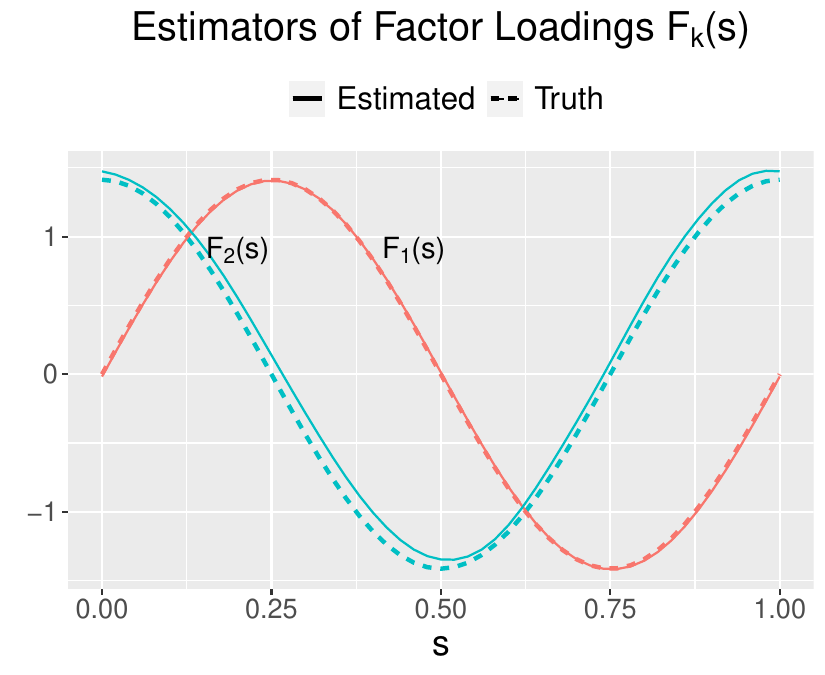}\hspace{-.2cm}
\includegraphics[scale=.41]{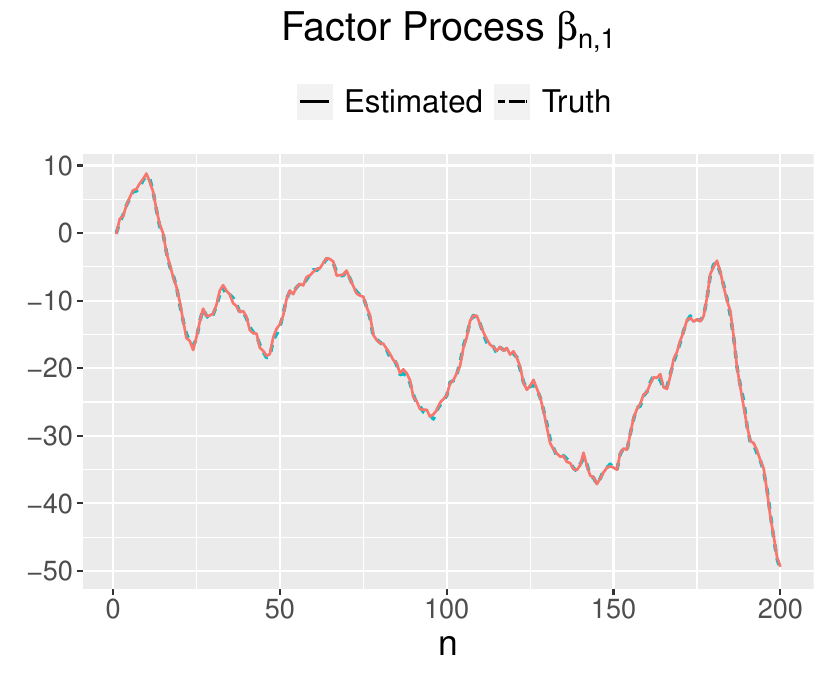}\hspace{-.2cm}
\includegraphics[scale=.41]{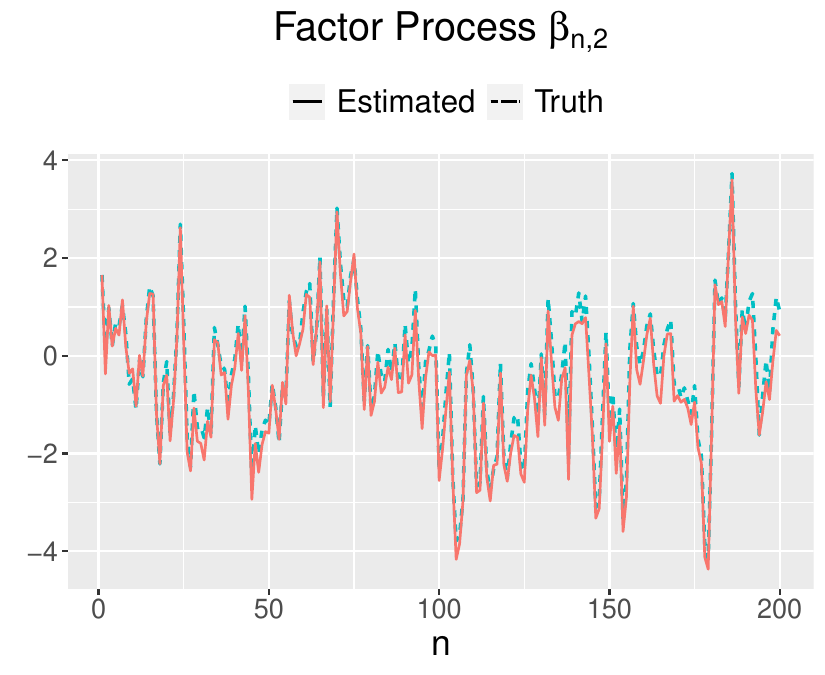}
%\vspace{-.5cm}
\caption{Estimators of $F_{1}(s)$ and $F_{2}(s)$ for the nonstationary FDF model and the estimated trajectory for the factor time series $\{\beta_{n,1}\}$ and $\{\beta_{n,2}\}$. In this example, $K=2$, $r=1$, and the sample size is $N=200$.} \label{ExampleEstimatorsNS}
\end{center}
\end{figure}
\begin{example} We simulate $\{X_{n}\}$ from the FDF model with $K=2$, $r=1$, and sample size $N=200$, where $F_{1}(s)=\sin(2\pi s)$, $F_{2}(s)=\cos(2\pi s)$, $\Delta \beta_{n,1}= 0.7 \Delta \beta_{n-1, 1} + u_{n,1}$, and $\beta_{n,2}= 0.7 \beta_{n-1, 2} + u_{n,2}$, with $s \in [0,1]$. Figure \ref{ExampleEstimatorsNS} shows the estimators obtained from Algorithm \ref{A2}. The left panel shows the factor loadings and their corresponding estimators, while the center and the right panels show the trajectory estimated for the factors. Similar to Example \ref{Example1}, we observe that the proposed estimators have good performance in this nonstationary example. A more exhaustive simulation study is performed in Section \ref{Simulation}.
\end{example}

\begin{remark} As we mentioned before, the estimators of the FDF model  are not unique, but the subspace $\mathcal{H}_{F}$ generated by the factor loadings is unique. Here, we defined $\widehat{F}_{k}$ as the eigenfunctions of the corresponding long-run covariance operator. However, users can choose an appropriate orthonormal rotation to obtain factors that allow a meaningful-interpretation. For example, one possible rotation is the well-known VARIMAX rotation.
%Thus, other basis functions of such subspaces can be considered.
\end{remark}

The dynamic modeling for the FDF model is attained by modeling the factor processes $\{ \widehat{\beta}_{n,k}\}$ and using the relationship $\widehat{X}_{n}(s)= \sum_{k=1}^{\widehat{K}} \widehat{\beta}_{n,k} \widehat{F}_{k} (s)$.

In general, the operator $\Lambda$ defined in \eqref{Op_Lambda} represents the loading space and the dynamic of the functional time series. Moreover, if the functional time series is an $I(1)$ functional process, then the operator $\Lambda$ classifies the space of common trends and the cointegrating space. We conclude that the dynamics of the FDF model over time are accurately described by using the space generated by this operator.

\section{Estimation properties}\label{Properties}
In this section, we study the properties of the estimators described in Section \ref{FDFM}. For this, we assume that  $X_{n}$ has a functional linear representation in $L^{2}_{\mathcal{H}}$. Without loss of generality, we assume that the curves are defined on the unit interval $D=[0,1]$ and $\mathbb{E}(X_{n})=0$. 
\begin{assumption}\label{P1}
The functional time series $\{X_{n}\}$ has a linear representation $X_{n} = \sum_{j=0}^{\infty} \Psi_{j}(\epsilon_{n})$, with $\Psi_{j}\in \mathcal{B}_{\mathcal{H}}$, and $\sum_{m=1}^{\infty}\sum_{j=m}^{\infty} \|\Psi_{j}\|<\infty$.
\end{assumption}		
\begin{assumption}\label{P2}
$\int \mathbb{E} \{X_{0}^{2} \}(s)\mathrm{d}s<\infty $ and $\lim_{m\to \infty}m\left( \mathbb{E}\left[\int\{X_{n}(s)- X_{n,m}(s) \}^{2}\mathrm{d}s\right]\right)^{1/2}=0 ,$ where $X_{n,m}= \sum_{j=0}^{m-1} \Psi_{j}(\epsilon_{n})+\sum_{j=m}^{\infty} \Psi_{j}(\epsilon_{n}^{(m)})$, and for each $m$, $\{\epsilon_{k}^{(m)}\}$ is an  independent copy of $\{\epsilon_{k}\}$.
\end{assumption}		

Assumption \ref{P1} does not impose any restrictions on the model \ref{Funct-Model}. For example,  if the factors processes are (scalar) linear processes, $\beta_{n,k}= \sum_{j=0}^{\infty} \theta_{j,k} u_{n-j,k}$ with $\{u_{n}\}$ an i.i.d. sequence, then $X_{n}(s)= \sum_{k=1}^{K} \sum_{j=0}^{\infty} \theta_{j,k} u_{n-j,k}F_{k}(s) + \varepsilon_{n}(s)$. This latter expression can be rewritten as $X_{n}(s)= \sum_{j=0}^{\infty} \Psi_{j}(\epsilon_{n})(s)$, with $\epsilon_{n}:= \sum_{k=1}^{K} u_{n,k}F_{k}+  \varepsilon_{n}(s)$ and $\Psi_{j}:= \sum_{k=1}^{K} \theta_{j,k} F_{k}\otimes F_{k}$. In this case, the condition $\sum_{m=1}^{\infty}\sum_{j=m}^{\infty} \|\Psi_{j}\|<\infty$ is equivalent to $\sum_{j=0}^{\infty} j |\theta_{j,k} |<\infty $. This latter condition is a common assumption on scalar time series in order to obtain basic results.

\begin{proposition} \label{Prop:S1}
Let $\{X_{n}, \, n\in \mathbb{Z} \}$ be a functional time series following model \eqref{Funct-Model}, and let $\widehat{\Lambda}$ be the estimator of $\Lambda$ described in Section \ref{FDFM}, with kernels $\widehat{g}(s,t)$ and $g(s,t)$, respectively. Under Assumptions \ref{A1}, \ref{A3}, \ref{P1}, and \ref{P2}, 
$\int \int \{\widehat{g}(s,t)- g(s,t)\}^{2} \mathrm{d}s\mathrm{d}t  \overset{\mathbb{P}}{\to} 0,$ as $N \to \infty$ and $p\to \infty$.
\end{proposition}

Proof: See Appendix.

Assumption \ref{P1} implies that $\{X_{n}\}$ is a stationary  functional time series. Thus, Proposition \ref{Prop:S1} corresponds to the stationary case of model \eqref{Funct-Model}.

As we noted above, the factors $\{\beta_{n,k}\}$ and factor loadings $F_{k}$ are not uniquely determined in the FDF model. However, the factor loading space is uniquely determined. We only study the theoretical properties of the factor loading space under the assumption that $K$ is known. Here, we say that a subspace $\mathcal{H}_{n}$ converges to a subspace $\mathcal{H}_{0}$ if $\|\Pi_{n} - \Pi_{0} \| \to 0$, where $\Pi_{n}$ and $\Pi_{0}$ are the orthonormal projectors on $\mathcal{H}_{n}$ and  $\mathcal{H}_{0}$, respectively, and $\|\cdot \|$ an operator norm. We use the notation $\mathcal{H}_{n}\to \mathcal{H}_{0}$ to indicate this convergence of a subspace.

%We note that the loading space $\mathcal{H}_{F}$ is equal to the space $\mathrm{Im}\, (\Gamma - \Gamma_{0})$. Thus, we have the following corollary.

\begin{corollary}\label{C1}  Let $\{X_{n}, \, n\in \mathbb{Z} \}$ be a functional time series following model \eqref{Funct-Model} with $K$ known. Suppose that $\Gamma_{0}$ is invertible. Then, under Assumptions \ref{A1}, \ref{A3}, \ref{P1} and \ref{P2},
$$\widehat{\mathcal{H}}_{F}\overset{\mathbb{P}}{\to} \mathcal{H}_{F}.$$
\end{corollary}

To study the nonstationary case, we assume that $\Delta X_{n}$ admits a functional linear representation as well.

\begin{assumption}\label{NonS}  The functional time series $\{X_{n}, \, n\in \mathbb{Z} \}\in L^{2}_{\mathcal{H}}$ is an $I(1)$ functional process so that $\{ \Delta X_{n}, \, n \geq 1\}$ is a stationary process and admits the representation 
\begin{equation}\label{DiffLinear}
\Delta X_{n} = \sum_{j=0}^{\infty} \Phi_{j} (\epsilon_{n-j}), \quad n\geq 1,
\end{equation}
with $ \Phi_{j} \in \mathcal{B}_{\mathcal{H}}$, $ \Phi_{j_{0}} \neq 0$ for some $j_{0}>0$, and the covariance operator of $\epsilon_{0}$, $\Gamma_{\epsilon_{0}}$, is positive definite. Furthermore,  Assumptions \ref{P1} and \ref{P2} are satisfied on the functional process $\{ \Delta X_{n}, \, n \geq 1\}$.
\end{assumption}

Assumption \ref{NonS} does not impose any restrictions on the model \eqref{Funct-Model}. If any factor process $\{\beta_{n,k}, \, \in \mathbb{Z}\}$ is an $I(1)$ process, then the functional time series that follows model \eqref{Funct-Model} can always be written as \eqref{DiffLinear}, with $ \Phi_{j}$ compact and self-adjoint operators. The condition $ \Phi_{j_{0}} \neq 0$ for some $j_{0}>0$ guarantees that $\Delta X_{n}$ is not a functional white noise. Consequently, $X_{n}$ admits a solution of the form $X_{n}= X_{0} + \Phi \left(\sum_{i=1}^{n} \varepsilon_{n} \right) + \nu_{n} $, where $ \nu_{n} $ is a stationary functional time series, $\{\varepsilon_{n}\}$ is an i.i.d. sequence, and $\Phi= \sum_{j\geq 0} \Phi_{j}$. 

From Proposition \ref{Prop:S1}, we have that $\widehat{\Lambda}_{\Delta X}$ is a consistent estimator of $\Lambda_{\Delta X}$.

\begin{proposition}\label{Prop:CS}
Let $\{X_{n}, \, n\in \mathbb{Z} \}$ be a nonstationary functional time series such that it follows model \eqref{Funct-Model}. Then, under Assumptions \ref{A1} and \ref{NonS} we have
that $\mathcal{H}_{F}^N =\{\mathrm{ker}( \Lambda_{\Delta X})\}^{\bot}$, and   $\mathcal{H}_{F}^S= \mathrm{ker}( \Lambda_{\Delta X})$.
\end{proposition}
Proof: See Appendix.

Proposition \ref{Prop:CS} shows that the $I(1)$ dynamics of the factors are recovered in the subspace generated by nonzero elements of $ \mathrm{Im} \, \Lambda_{\Delta X}$.

\begin{corollary} \label{C2} Let $\Gamma_{0}$ be invertible, and $K$ and $r$ known. If the assumptions in Proposition \ref{Prop:CS} hold, then $\widehat{\mathcal{H}}_{F}^S \overset{\mathbb{P}}{\to} \mathcal{H}_{F}^S$ and $\widehat{\mathcal{H}}_{F}^N\overset{\mathbb{P}}{\to} \mathcal{H}_{F}^N$.
\end{corollary}

\begin{corollary}  \label{C3}
Let $\{\widehat{\beta}_{n,k}\}$ be the estimated trajectory obtained from Algorithm \ref{A2} with $r$ and $K$ known. Then, if the assumptions in Proposition \ref{Prop:CS} hold, we have that, for $k=1,\ldots, r$, $\{\widehat{\beta}_{n,k},  n\in \mathbb{Z}\}$ is an $I(1)$ process, and if $r<K$, for $k=r+1, \ldots, K$, $\{\widehat{\beta}_{n,k},  n\in \mathbb{Z}\}$ is an $I(0)$ process.
\end{corollary}

Corollaries \ref{C2} and \ref{C3} show that the loading space is consistently estimated for both stationary and nonstationary components.

\section{Simulation study}\label{Simulation}
We study the performance of our proposed estimators for the FDF model. We compare our results with estimators obtained from functional PCA. To obtain the functional PCA estimators, we replace $\widehat{F}_{k}$ with $\widehat{F}_{k}^{\mathrm{PCA}}=\widehat{\varsigma}_k$ in Algorithm \ref{A1}, where $\widehat{\varsigma}_k$ denotes the $k$th eigenfunction of $\widehat{\Gamma}_{0}$ corresponding to the $k$th largest eigenvalue \citep[][]{Dominik}. Similarly, in Algorithm \ref{A2}, we consider the covariance operators at lag zero of $\{\Delta X_{n}\}$ and $\{Z_{n}\}$ instead of $\widehat{\Lambda}_{\Delta X}$ and $\widehat{\Lambda}_{Z}$. The PCA estimators are known to perform well when observations are uncorrelated.

\subsection{\textit{Simulation setting}}
We simulate $\{X_n(s);\, s\in [0,1],\, n=1\ldots,N\}$ from the FDF model with four different models defined as follows:
\begin{enumerate}
\item\label{Model1} Model 1: $K=1$, $F_{1}(s)= \sin(2\pi s )$, and $\beta_{n,1} = 0.7 \beta_{n-1} + u_{n,1}$.
\item\label{Model2} Model 2: $K=2$, $F_{1}(s)= \sin(2\pi s )$, $F_{2}(s)= \cos(2\pi s )$, $\beta_{n,1} = 0.8 \beta_{n-1,1} + u_{n,1}$, and $\beta_{n,2}=-0.5 \beta_{n-1,2}+ u_{n,2}$.
\item\label{Model3} Model 3: $K=1$, $r=K$, $F_{1}(s)= \sin(2\pi s )$, and $\Delta \beta_{n,1}= 0.5 \beta_{n-1,1} + u_{n,1}$.
\item\label{Model4} Model 4: $K=2$, $r=1$, $F_{1}(s)= \sin(2\pi s )$, $F_{2}(s)= \cos(2\pi s )$, $\Delta \beta_{n,1}= 0.7 \beta_{n-1,1} + u_{n,1}$, and $\beta_{n,2}= 0.5 \beta_{n-1,2} + u_{n,2}$.
\end{enumerate}
The processes $\{u_{n,i}\}$ are scalar white noises. In all cases, the functional white noise $\varepsilon_{n}(s)$ in model \ref{Funct-Model} is simulated as $\varepsilon_{n}(s)=W_{n}(s)$, where $W_{n}(s)$ is a Brownian motion in $[0,1]$. Models \ref{Model1} and \ref{Model2} represent stationary FDF models, while Models \ref{Model3} and \ref{Model4} represent nonstationary FDF models.

We evaluate the performance by computing the Integrated Squared Error (ISE) value defined as:
$$\mathrm{ISE}_{F}= \int_{0}^{1}\! \left\{F(s) - \widehat{F}(s) \right\}^2 \mathrm{d}s.$$
We consider sample sizes $N=200, 300, 500$, and $1000$. We simulate $1000$ replications for each model. For each replication, we estimate the factor loadings and factor processes. Then, we compute the error value $\mathrm{ISE}_{F}$, both for our proposed estimators and for PCA estimators.

Additionally, we report the estimated factor number $\widehat{K}$ for each simulation using \eqref{EstNumF}. % and scree plot.

\subsection{\textit{Simulation results}}
\begin{figure}[!b]
\begin{center}
\includegraphics[scale=.5]{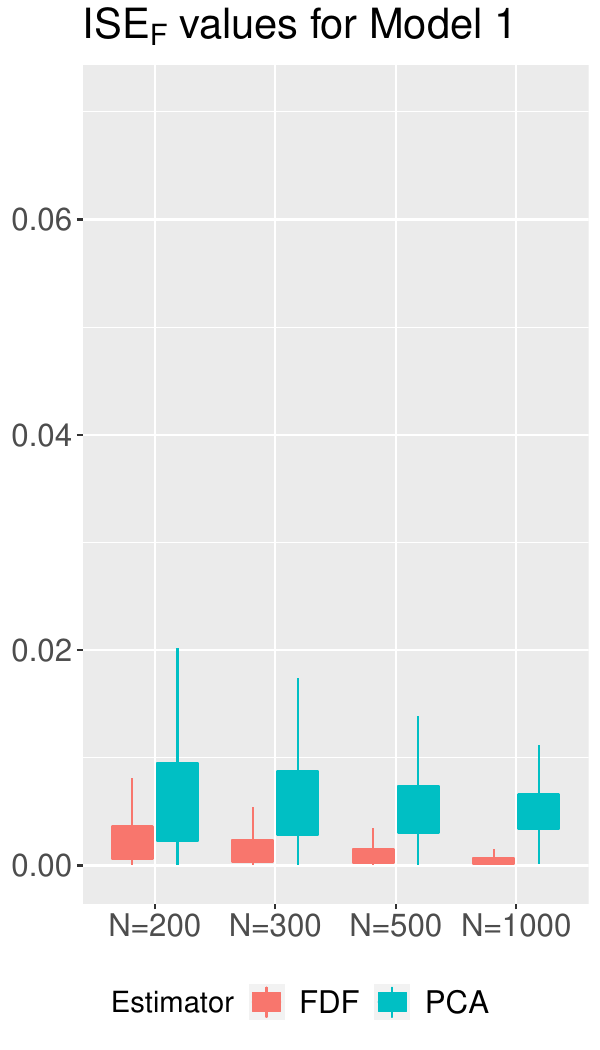}
\includegraphics[scale=.5]{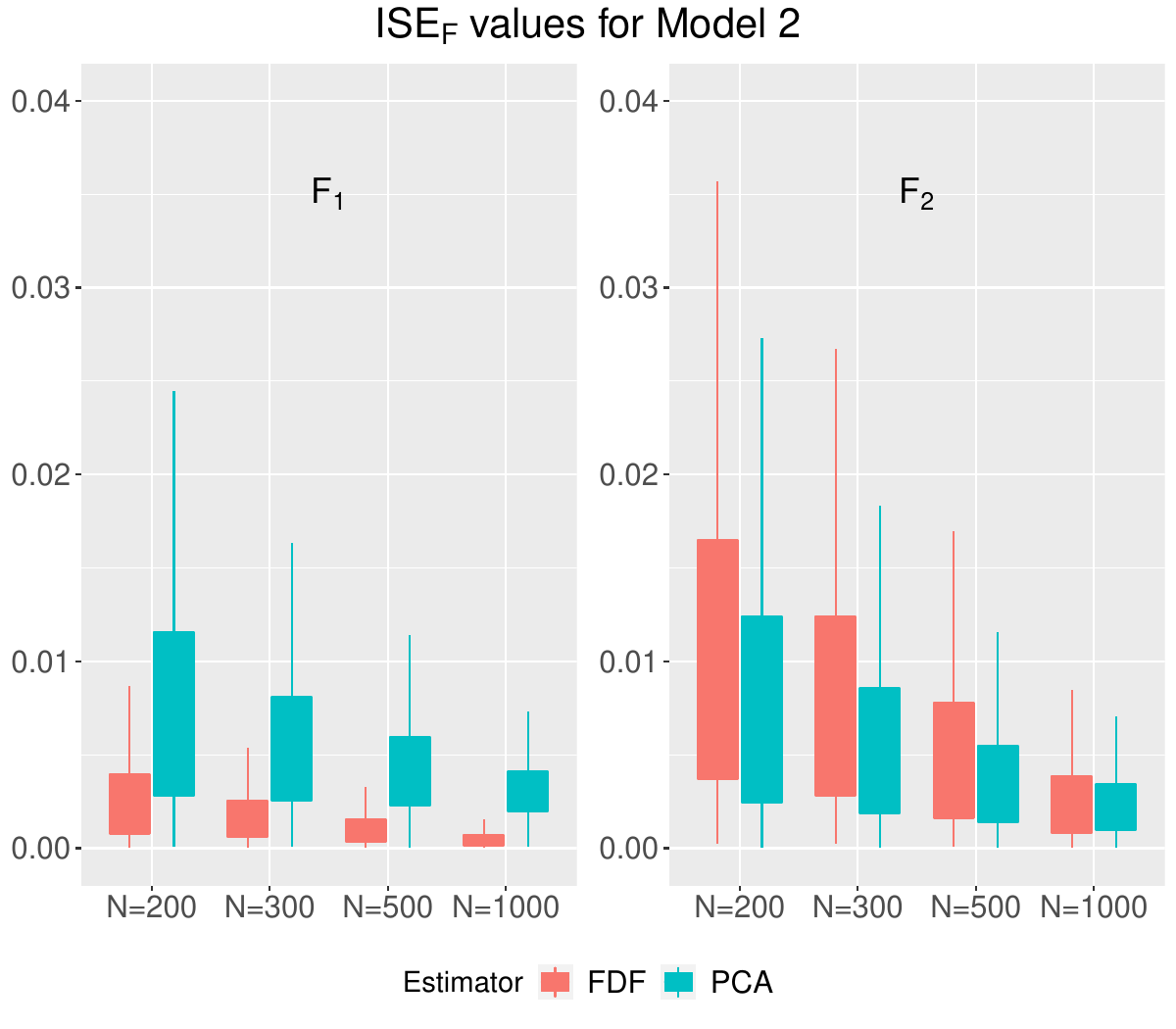}
\includegraphics[scale=.5]{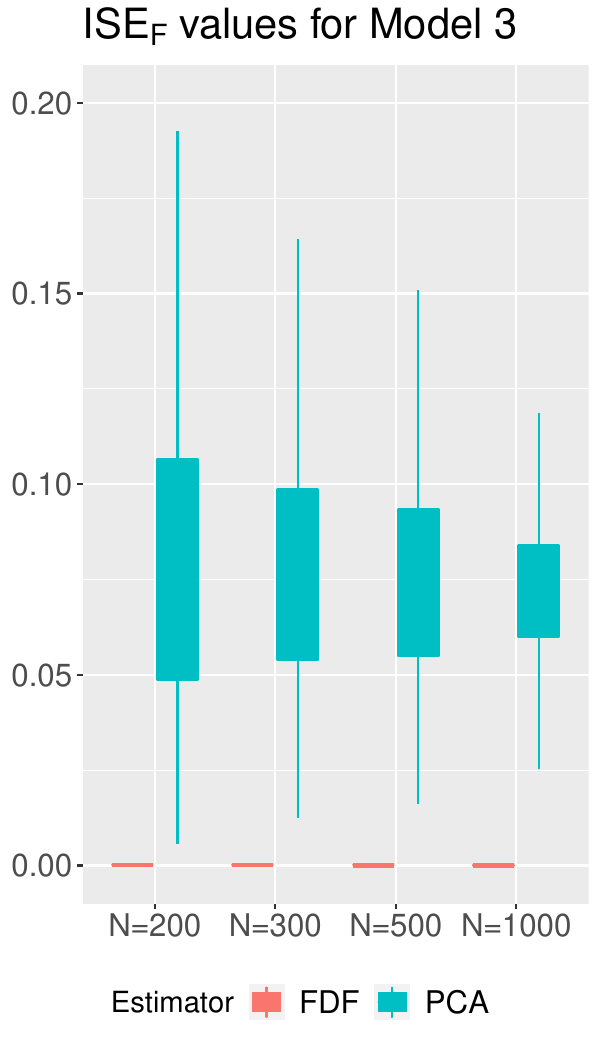}
\includegraphics[scale=.5]{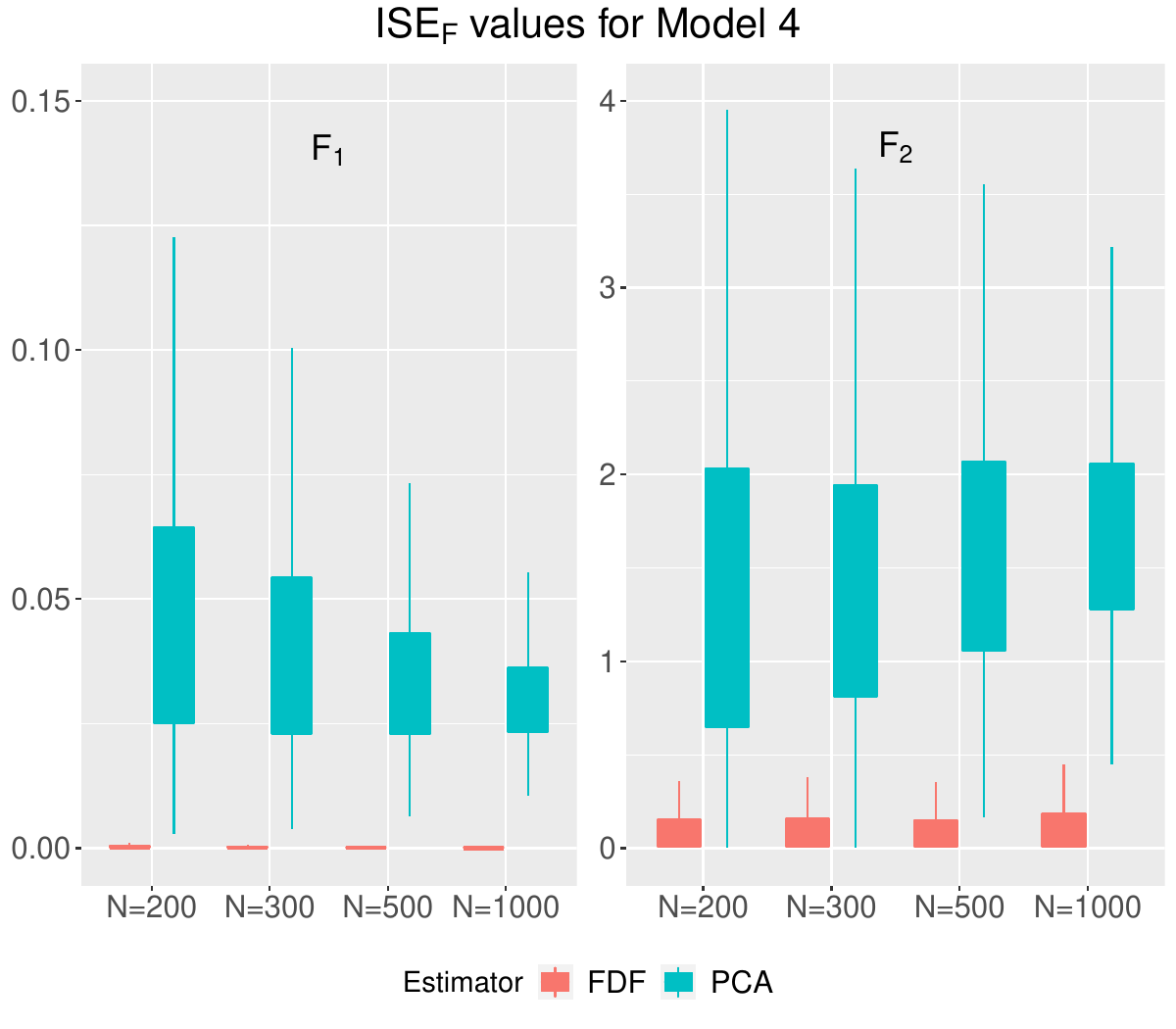}
\caption{Boxplots of $\mathrm{ISE}_{F}$ values for each simulation. In all cases, our proposed estimators (in red) outperform the PCA estimators. Although for $F_{2}$ in Model \ref{Model2}, the PCA estimator performs as well as our estimator.} \label{FigErrors1}
\end{center}	
\end{figure}

We denote by FDF the simulation results when considering our nonparametric estimators. First, we describe the results for $\mathrm{ISE}_{F}$ values. In doing so, we assume that the true number of factors is known.
Figure \ref{FigErrors1} shows boxplots of the results. Each plot presents the $\mathrm{ISE}_{F}$ values for a specific $F_{i}$ with different sample sizes. For Models \ref{Model1} and \ref{Model3}, we have only one loading factor, and for Models \ref{Model2} and \ref{Model4}, we have two loading factors, $F_{1}$ and $F_{2}$.

For Model \ref{Model1} (Figure \ref{FigErrors1} top left), we observe that our proposed estimator is highly accurate and presents lower error values than the PCA estimator. These error values decrease when the sample size increases. For Model \ref{Model2}, we observe that for the first loading factor $F_{1}$ (Figure \ref{FigErrors1} top center), the results are similar to the results of Model \ref{Model1}. Our method outperforms the PCA estimator. For $F_{2}$ (Figure \ref{FigErrors1} top right), we see that the PCA method performs as well as our method. Thus, we conclude that, in general, our method outperforms the PCA estimator for the stationary Models \ref{Model1} and~\ref{Model2}.

Now, we analyze the nonstationary models (Models \ref{Model3} and  \ref{Model4}). We observe that in all cases, our method has the lowest error values. The $\mathrm{ISE}_{F}$ values for our method remain as accurate as in the stationary cases. However, the $\mathrm{ISE}_{F}$ values for the PCA method become significantly larger. We conclude that our proposed method performs well in all simulated cases and outperforms the PCA method.
\begin{figure}[!b]
\begin{center}
\includegraphics[scale=.325]{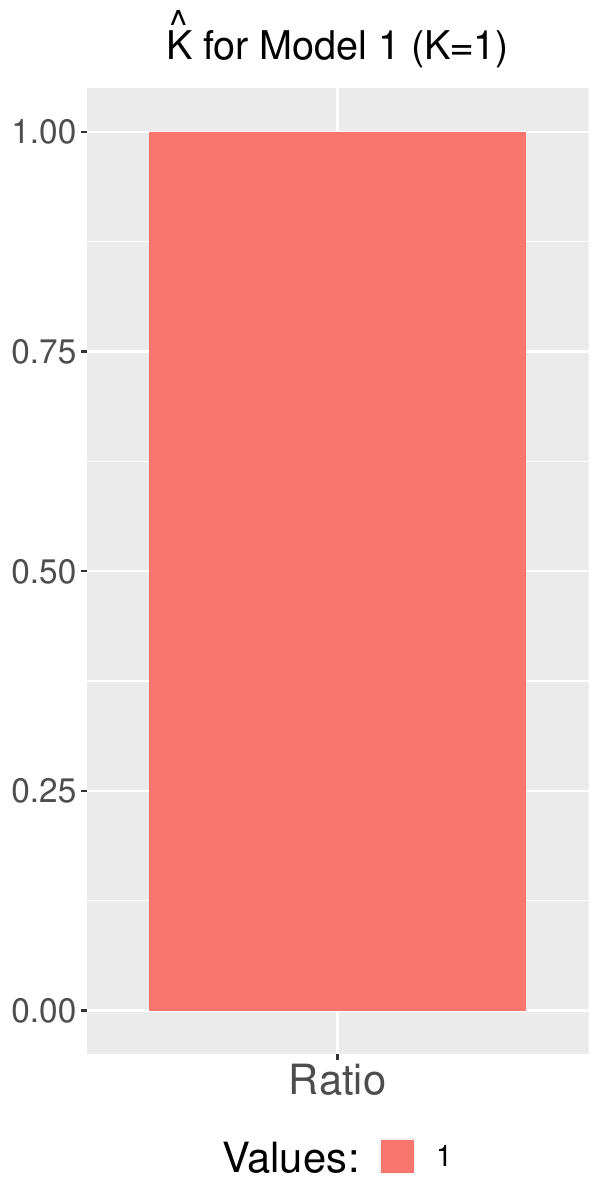}
\includegraphics[scale=.325]{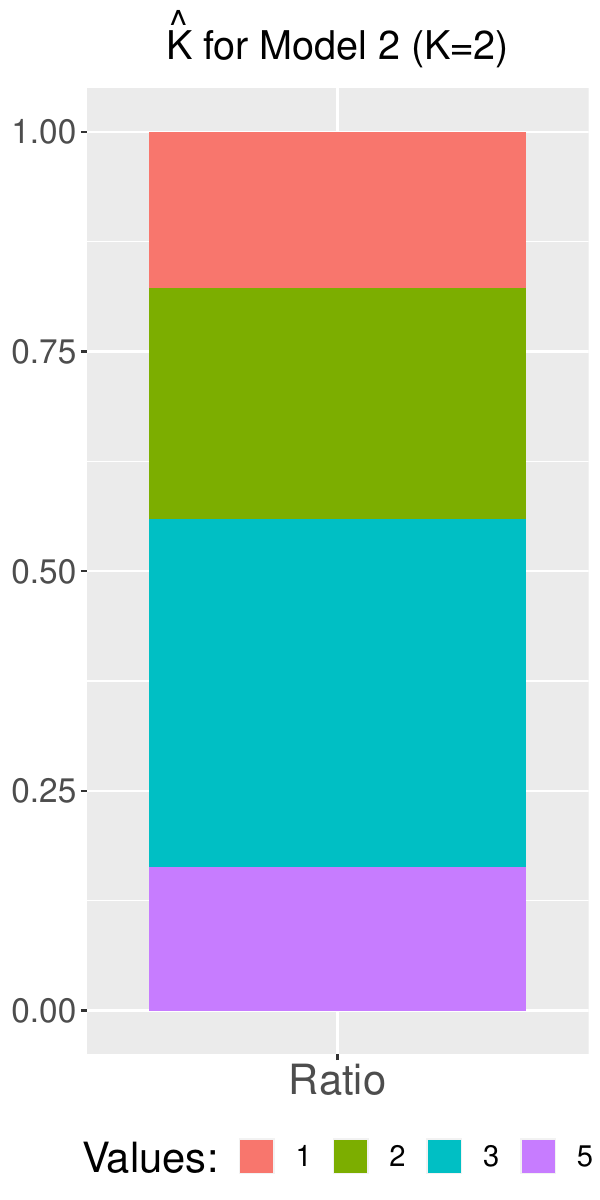}
\includegraphics[scale=.325]{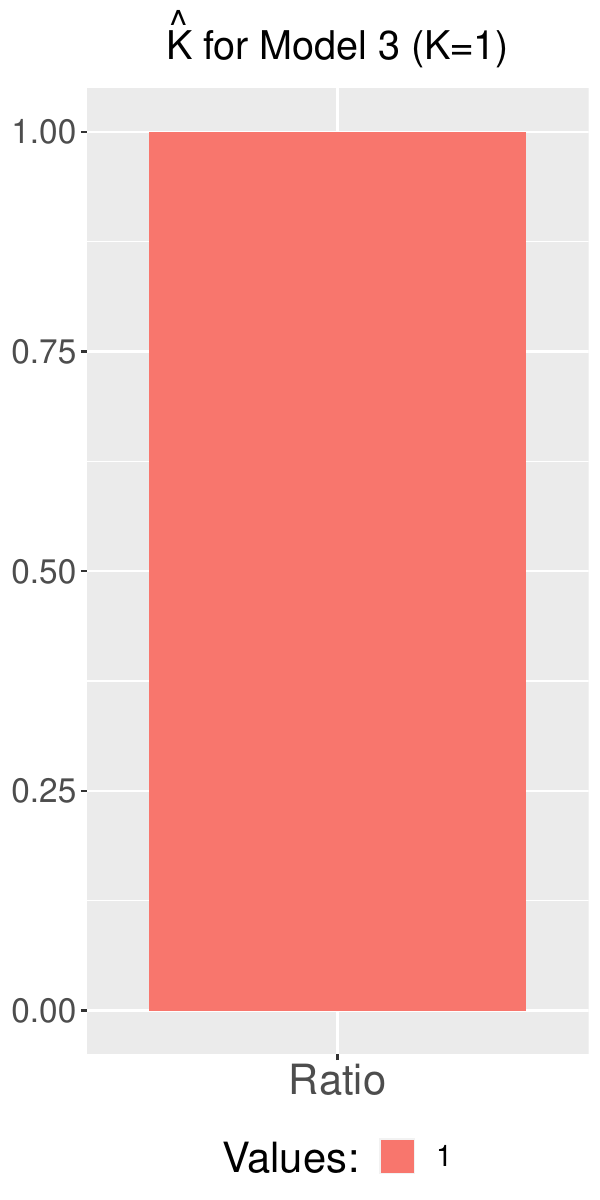}
\includegraphics[scale=.325]{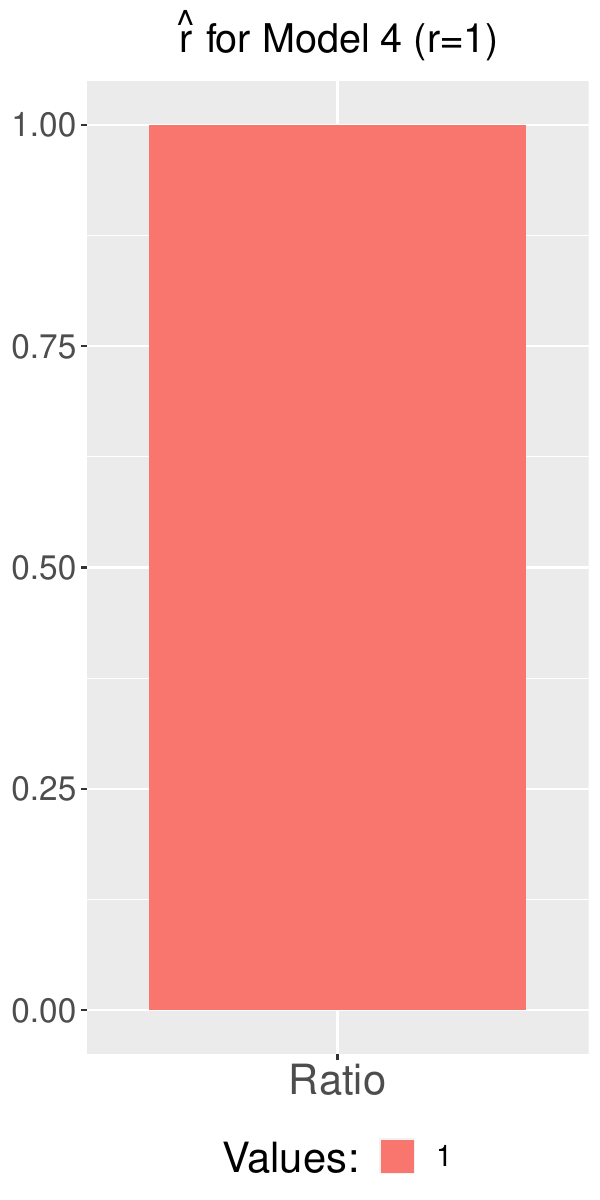}
\includegraphics[scale=.325]{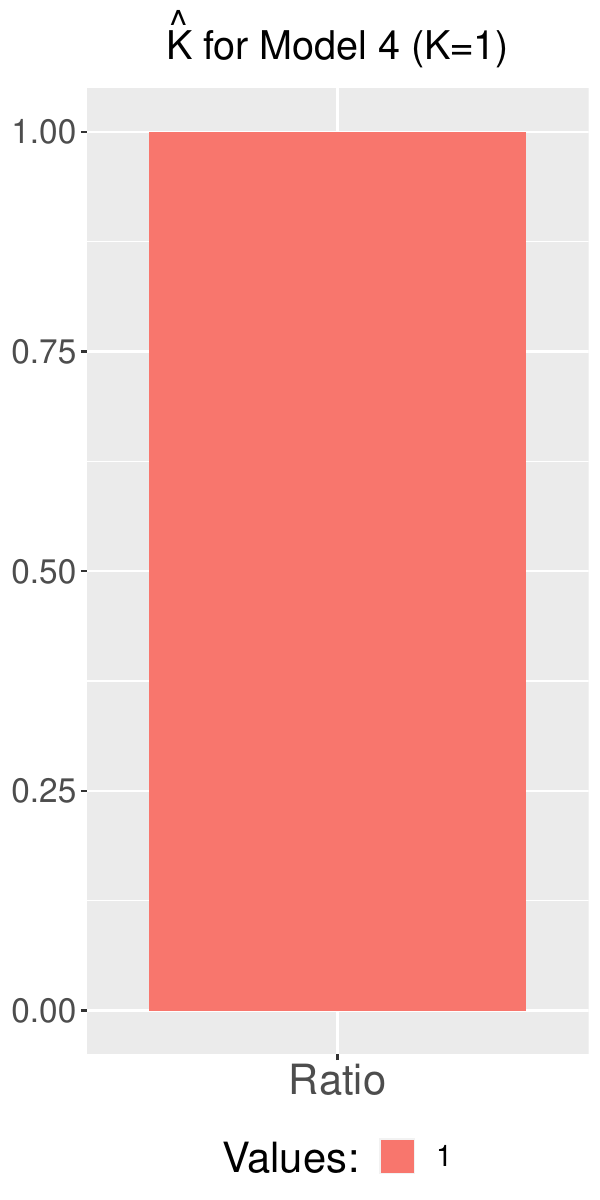}
\caption{Values and proportions of the number of factors estimated for sample size $N=300$, and $1000$ replications. The ratio method provides a good estimator for $K$.} \label{FigHatK}
\end{center}
\end{figure}

Finally, we show the results corresponding to the estimation of the number of factors. For each replication, we estimate the number of factors using \eqref{EstNumF}. We only present the results from sample size $N=300$, since the results from the sample sizes $N=200,500,$ and $1000$ are similar.
Figure \ref{FigHatK} shows the estimated values $\widehat{K}$ over the 1000 replications. Each bar uses colors to represent the proportion of replications. %where $\widehat{K}=i$, $i=1,2,3,\ldots,$ depending on the criteria.

For Model \ref{Model1}, we observe that with the ratio method, we obtain $\widehat{K}=1$ for all replications. 
For this Model \ref{Model1}, we conclude that the ratio method correctly estimates the value of $K$.
For Model \ref{Model2}, $\widehat{K}$ takes values in $\{1,2,3,5\}$, resulting in $\widehat{K}=2$ and $\widehat{K}=3$ with a large proportion. For Model \ref{Model3}, the ratio method successfully estimates the value of $K$. For Model \ref{Model4}, we need to estimate $r$ and $K-r$, where $r$ is the number of nonstationary factor processes and $K-r$ is the number of stationary factor processes. In this case, we observe that the ratio method provides the correct values of $r$ and $K-r$. 

In general, the ratio method correctly estimates the number of factors with the exception of Model \ref{Model2}. 

We conclude that our proposed methodology performs well and is superior to the PCA method under time dependence.

\section{Data application}\label{Application}
\section{Data application}\label{Application}
We fit the FDF model with our proposed nonparametric estimators to analyze the Federal Reserve interest rates. Then, we study the estimated factor loadings and the trajectories of the factor processes.

\subsection{\textit{Yield curve}}
\begin{figure}[!t]
\begin{center}
\includegraphics[scale=.41]{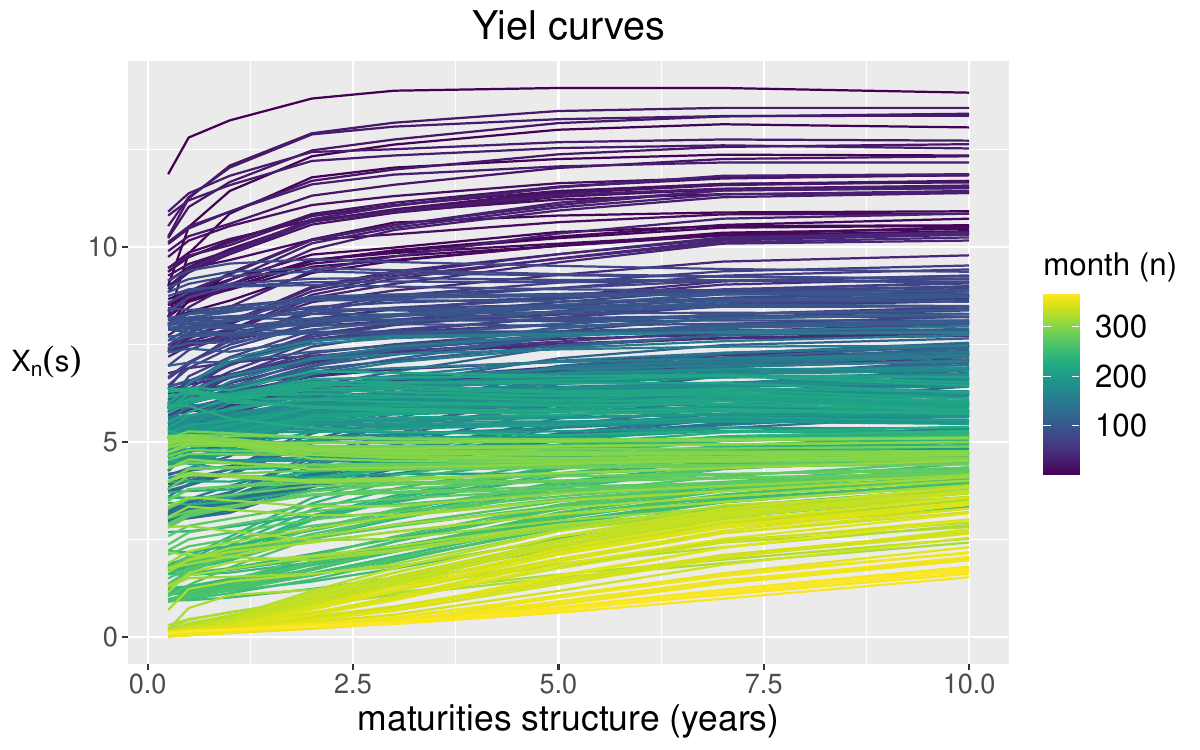}
\includegraphics[scale=.41]{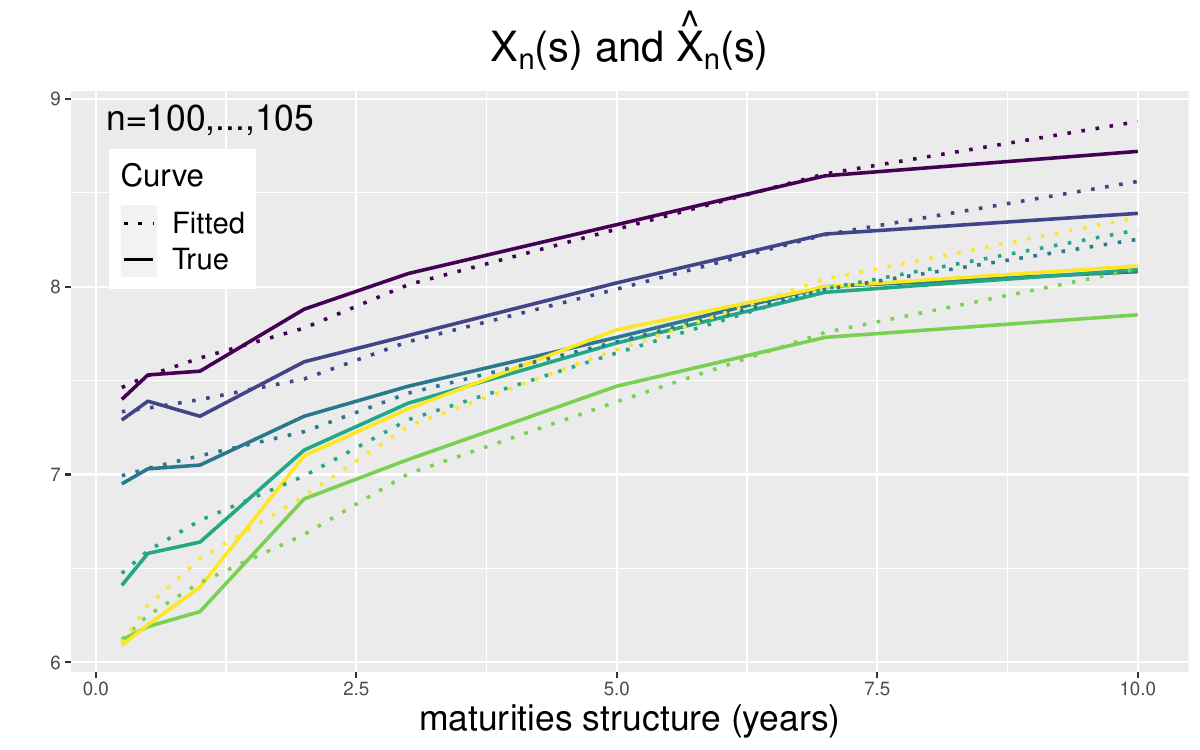}
\caption{ Right: Monthly curves of yield data from June $1982$ to May $ 2012$. Left: six consecutive yield curves from the dataset (solid curves) and the corresponding five yield curves obtained from the FDF model with the estimators (dashed curves), that is, $\widehat{X}_{n}(s) = \sum_{k=1}^{3} \widehat{F}_{k} (s)\widehat{\beta}_{n,k}$ } \label{data1}
\end{center}
\end{figure}

The interest rate data from the Federal Reserve are available in the R package \textit{YieldCurve}. The dataset represents monthly yield data from June $1982$ to May $ 2012$ at different maturities: $3$ months, $6$ months, $1$ year, $2$ years, $3$ years, $5$ years, $7$ years and $10$ years (see Figure \ref{data1}). Yield curves are important in economics and finance and can help to determine the current and future position of the economy. The Nelson-Siegel parametrization is commonly used to describe yield curves \citep[see][]{NelsonandSiegel87, Diebold2006, KoopmanEtAl2010}. Our proposal is a nonparametric estimator and can be considered as alternative modeling to the Nelson-Siegel modeling. To justify the functional approach, we assume that the dataset is continuous on maturities. That is, a curve $X_{n}(s)$ represents interest rates in month $n$ with time to maturity $s$. To estimate the continuous curves, we fit $15$ cubic B-spline basis functions for each monthly observation, i.e., $X_{n}(s)=\sum_{r=1}^{15} b_{r,n} \phi_{r}(s)$, where $\{\phi_{1}, \ldots, \phi_{15}\}$ is the B-spline basis function, with $n=1, \ldots, 366$.

Yield curves were analyzed with a nonparametric approach and with a functional approach. \cite{HaysSpencerShen2012} used a functional approach combined with an EM algorithm to jointly estimate the factor loading curves and the factors. Their approach is difficult to apply if more data are observed at different maturities and almost impossible to apply if the functional data are observed in a dense set. In contrast, our estimators are easy to implement and can be applied to either dense or sparse observations and for stationary and nonstationary functional time series.

\begin{figure}[!b]
\begin{center}
\includegraphics[scale=.38]{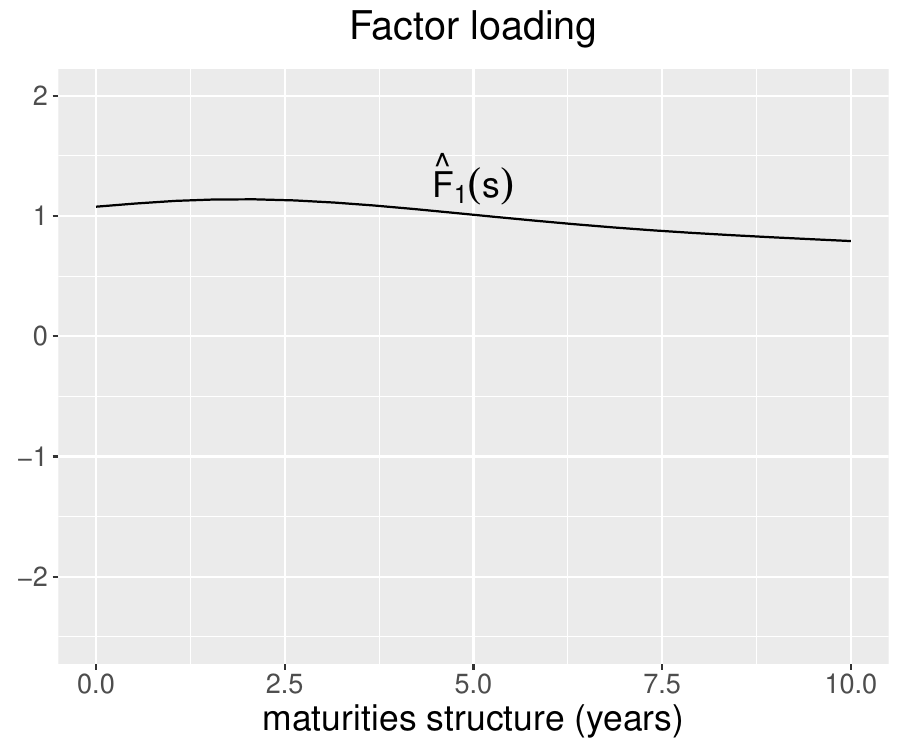}\hspace{-.2cm}
\includegraphics[scale=.38]{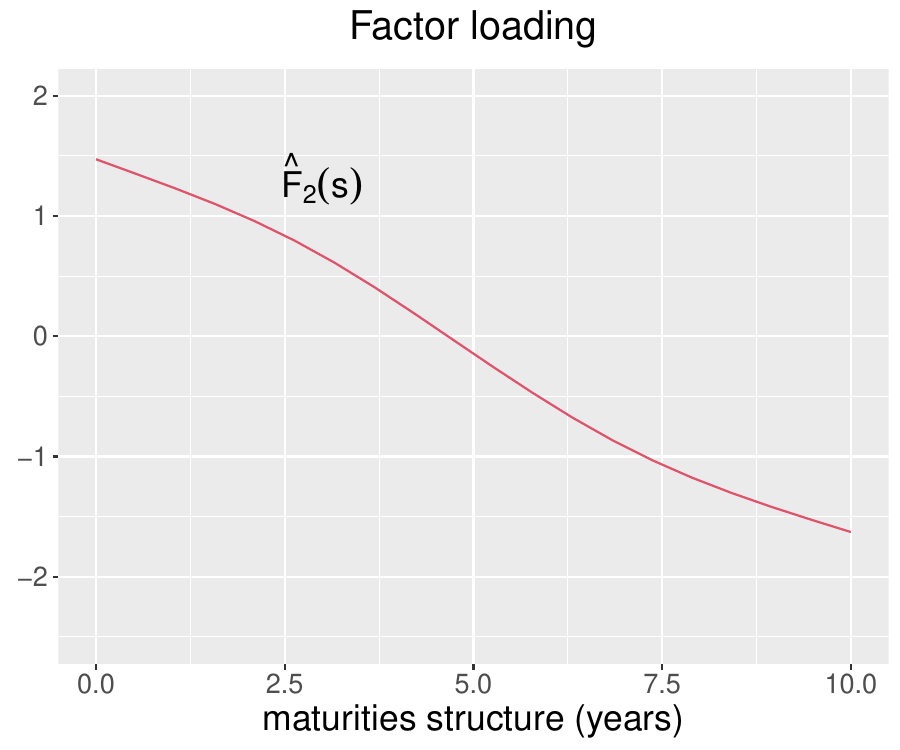}\hspace{-.2cm}
\includegraphics[scale=.38]{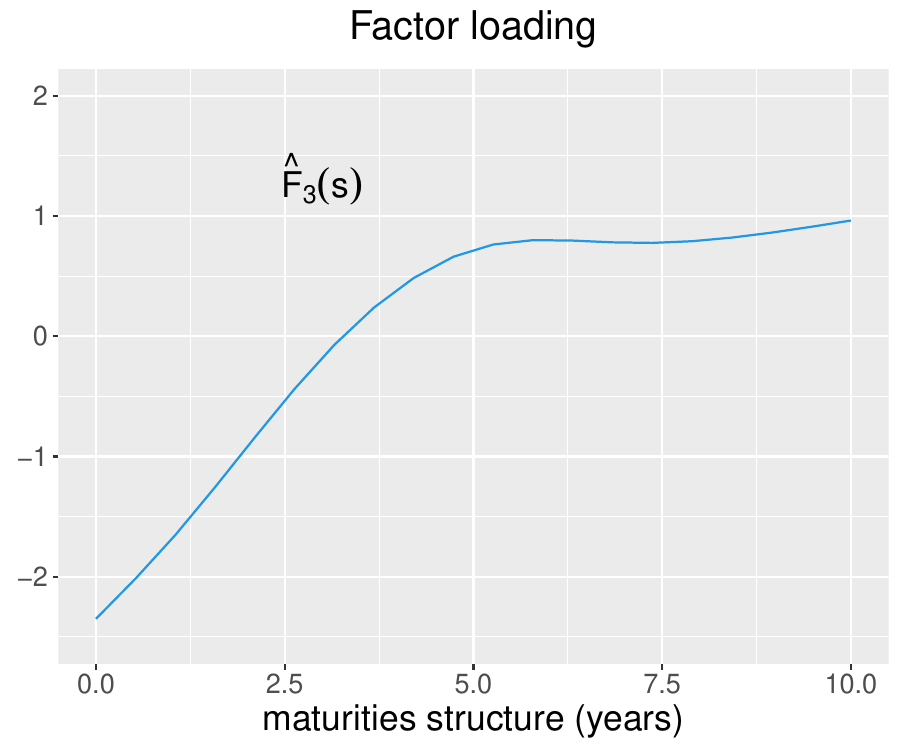}
\includegraphics[scale=.38]{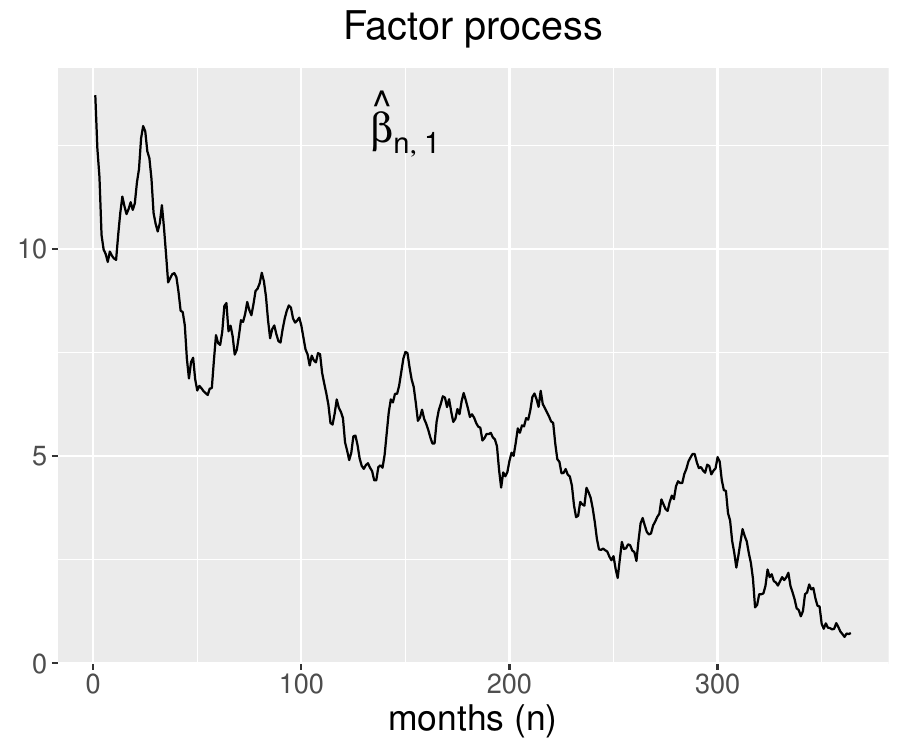}\hspace{-.2cm}
\includegraphics[scale=.38]{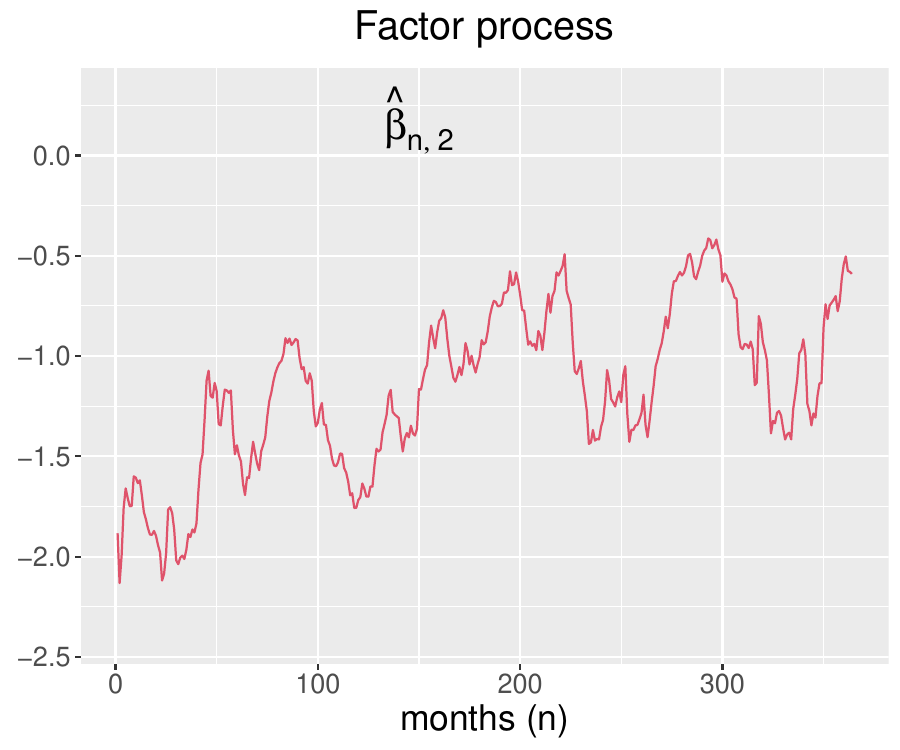}\hspace{-.2cm}
\includegraphics[scale=.38]{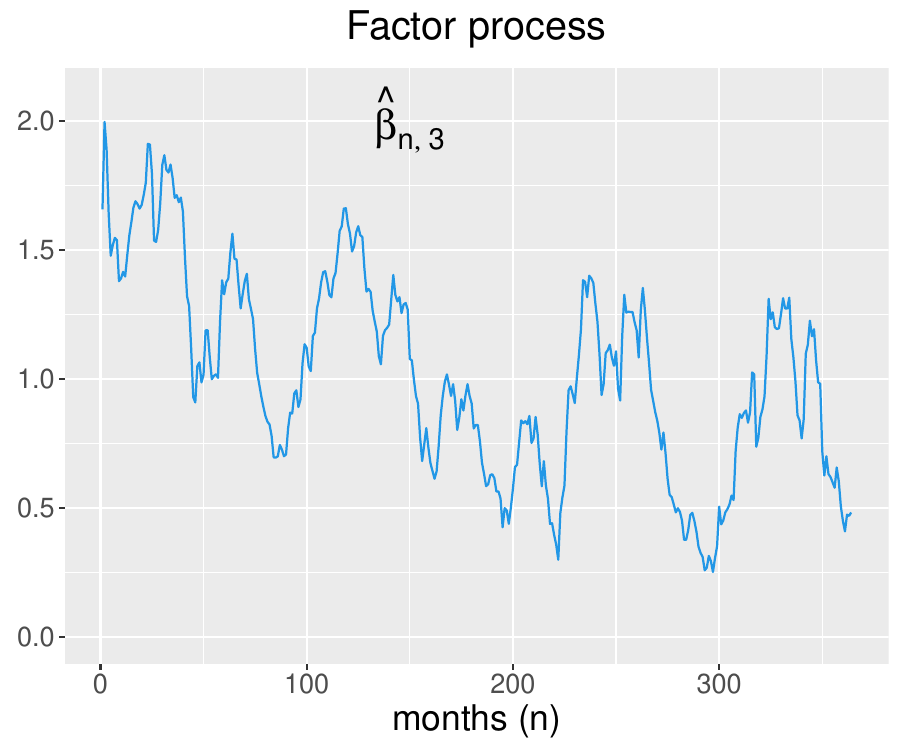}
\caption{Estimator of factor loading curves, $\widehat{F}_{1}, \widehat{F}_{2},$ and $\widehat{F}_{3}$, and the corresponding trajectories of the factor processes estimated, $\{\widehat{\beta}_{n,k}, 1\leq n \leq 366\}$. } \label{Estimators}
\end{center}
\end{figure}

We are interested in studying the factor loading curves and the factor processes that drive the interest rates at different maturities by taking into account the dependence structure of the functional time series. To infer the stationarity of the functional time series, we apply a test proposed by \cite{HorvathKokoszkaetal2014}. The $p$-value of the test, with $15$ basis functions, is equal to $0.001$, and the smaller the $p$-value, the more evidence there is against stationarity. Therefore, we follow Algorithm \ref{A2} with three factors, and $r=1$.

Figure \ref{Estimators} shows the results. In the first row, we plot the estimator of factor loading curves, $\widehat{F}_{1}, \widehat{F}_{2},$ and $\widehat{F}_{3}$, and in the second row, we plot the trajectories of the factor processes estimated, $\{\widehat{\beta}_{n,k}, 1\leq n \leq 366\}$. Since the factor loading curves are time-invariant, they represent the common properties of the yield curves, while the factor processes represent the dynamics over time. Our estimators agree with the level, slope and curvature functions of the Nelson-Siegel parameterization described by \cite{Diebold2006}, which is interesting since we do not assume any specific model (see Appendix for a plot of Nelson-Siegel curves). Therefore, the first factor loading curve $F_{1}$ is the level, and it is associated with the long-term factor; these dynamics are described by $\{\beta_{n,1} \}$. The factor loading $F_{2}$ is the slope, and it is associated with the short-term factor that is represented by the process $\{\beta_{n,2} \}$. Finally, the factor loading $F_{3}$ is the curvature that corresponds to the medium-term factor, and the factor $\{\beta_{n,3} \}$ describes such dynamics.

Figure \ref{data1} (right) shows six consecutive yield curves from the dataset with the corresponding six yield curves obtained from the FDF model with the estimators, that is, $\widehat{X}_{n}(s) = \sum_{k=1}^{3} \widehat{F}_{k} (s)\widehat{\beta}_{n,k}$. These six curves correspond to the months of October, November, December, January, February, and March of $1990$ and $1991$, respectively. We observe that the yield curves seem to be accurately represented by the FDF model.

\section{Discussion}\label{sec:dis}
We propose new nonparametric estimators of the functional dynamic factor model, taking into account the time dependence of the functional data. We use the long-run covariance operator to define a subspace of the continuous functions where the dynamics of the functional data are properly described by the factor processes. We have showed that the proposed estimators of the factor loading curves represent a subspace where factor processes describe the dependence of the functional time series, under both the stationarity and nonstationarity assumptions. We compared our proposed estimators with eigenfunctions of the covariance operator. From the simulation study, we conclude that our proposed estimators have better performance than PCA-based estimators.

From a mathematical point of view, factor loading curves can be considered part of basis functions for the Hilbert space. In principle, factor loading curves could be, for example, eigenfunctions, but eigenvalues do not take into account the time-dependent structure and therefore do not represent the dynamics over time. The ideas developed here can be extended to multivariate functional time series for studying the common factors among the different functional time series and the factors of each functional time series that are not shared. The extended model might be relevant in many applications, such as multi-economy yield curves.

\baselineskip=26pt
\section*{Appendix A. Proofs}\label{sec:appendix}

\noindent \textbf{Derivation of equation \eqref{LongRCovLP}:} 
Let $Y_{n}(s)=\sum_{j=0}^{\infty} A_{j}(\varepsilon_{n-j})(s)$ be a stationary functional linear process, with $ \sum_{j=0}^{\infty} \| A_{j} \|< \infty$. Let $A= \sum_{j=0}^{\infty} A_{j}$. First, we observe that for all $z\in \mathcal{H}$, the covariance operator at lag $h$ holds
\begin{align*}
\langle \Gamma_{h}(z), z \rangle = \mathbb{E}( \langle Y_{0},z\rangle  \langle Y_{h} , z \rangle ) &=\sum_{i=0}^{\infty} \sum_{j=0}^{\infty} \mathbb{E}\{  \langle A_{i}(\varepsilon_{-i}) ,z \rangle  \langle A_{j} (\varepsilon_{h-j}), z \rangle \}\\
&=\sum_{i=0}^{\infty} \sum_{j=0}^{\infty} 
 \mathbb{E}\{  \langle \varepsilon_{-i} , A_{i}^{*}(z) \rangle  \langle \varepsilon_{h-j}, A_{j}^{*} (z) \rangle \}.
\end{align*}
We note that $\mathbb{E}\{  \langle \varepsilon_{-i} , A_{i}^{*}(z) \rangle  \langle \varepsilon_{h-j}, A_{j}^{*} (z) \rangle \}= \langle \Gamma_{\varepsilon_{-i},\varepsilon_{h-j}}  A^{*}_{i}(z), A^{*}_{j}(z) \rangle$. Since $\{\varepsilon_{n} \}$ are i.i.d., we have that $\Gamma_{\varepsilon_{i},\varepsilon_{j}}=0 $ if $i\neq j$. Thus, $ \langle \Gamma_{\varepsilon_{-i},\varepsilon_{h-j}} ( A^{*}_{i}(z)), A^{*}_{j}(z) \rangle =  \langle \Gamma_{\varepsilon_{0}} ( A^{*}_{i}(z)), A^{*}_{i+h}(z) \rangle =  \langle A_{i+h} \Gamma_{\varepsilon_{0}}  A^{*}_{i}(z), (z) \rangle$. Substituting this into the above expression, we have
$$
\Gamma_{h} = \sum_{i=0}^{\infty}   A_{i+h} \Gamma_{\varepsilon_{0}}  A^{*}_{i}.
$$
Now, we compute the long-run covariance operator $\Lambda$. For all $z\in \mathcal{H}$, we have
\begin{align*}
\langle \Lambda(z), z \rangle= \sum_{h=-\infty}^{\infty} \langle \Gamma_{h}(z), z \rangle & =  \sum_{h=-\infty}^{\infty} \sum_{i=0}^{\infty} \langle   A_{i+h} \Gamma_{\varepsilon_{0}}  A^{*}_{i}(z), z \rangle \\
&= \sum_{i=0}^{\infty} \langle   \Gamma_{\varepsilon_{0}}  A^{*}_{i}(z), \sum_{h=-\infty}^{\infty}  A^{*}_{i+h} (z) \rangle,    
\end{align*}
since $A_{j}=0$ for all $j<0$, and changing the variable $h$ to $k=h+i$, we obtain that $\sum_{h=-\infty}^{\infty}  A^{*}_{i+h} (z)= \sum_{k=0}^{\infty}  A^{*}_{k} (z)$. That is, $ \langle \Lambda(z), z \rangle=  \langle  \Gamma_{\varepsilon_{0}}  A^{*} (z), A^{*}(z) \rangle= \langle  A \Gamma_{\varepsilon_{0}}  A^{*} (z), z \rangle$. This implies that the long-run covariance operator is $\Lambda= A \Gamma_{\varepsilon_{0}}  A^{*} $.

\hfill $\Box$
\medskip

All operators involved in the proof are integral operators, i.e., each operator is represented by a kernel function in $L^{2}([0,1]\times[0,1])$. Thus, the operator norm is defined as the usual norm in $L^{2}([0,1]\times[0,1])$, e.g., $\|\Lambda \|^{2}= \int\! \int \! c_{-0}^{2} (s,t) \mathrm{d}s\mathrm{d}t $.\medskip

\noindent \textbf{Proof of Proposition \ref{Prop:S1}:}  Let us first observe that under Assumptions \ref{P1} and \ref{P2}, the estimator $\widehat{\Gamma}$ is a consistent estimator of $\Gamma$. Explicitly, we have that  $\int \! \int \!\{\widehat{c}(s,t)- c(s,t) \}^{2}\mathrm{d}s\mathrm{d}t \overset{\mathbb{P}}{\to} 0.$ This is because with Assumptions \ref{P1} and \ref{P2} the functional time series $\{X_{n}\}$ is an $L^{2}\, m$-approximable (and hence sttaionary). The reader is referred to \cite{HorvathEtAl2012} for the details of this proof.

Similarly, we have that  $\int \! \int \! \left[ \widehat{\gamma}_{0}(s,t)- \mathbb{E} \{X_{1}(s)X_{1}(t) \} \right]^{2}\mathrm{d}s\mathrm{d}t \overset{\mathbb{P}}{\to} 0.$ That is, $\| \widehat{\Gamma}_{0} - \Gamma_{0}\| \overset{\mathbb{P}}{\to} 0$. Thus, 
\begin{align*}
\int \! \int \!\{\widehat{c}_{-0}(s,t)- c_{-0}(s,t) \}^{2}\mathrm{d}s\mathrm{d}t &= \int \! \int \!\{\widehat{c}(s,t)- \widehat{\gamma}_{0}(s,t) - c(s,t) + \widehat{\gamma}_{0}(s,t) \}^{2}\mathrm{d}s\mathrm{d}t \\
& \leq 2 \int \! \int \!\{\widehat{c}(s,t)- c(s,t) \}^{2}\mathrm{d}s\mathrm{d}t + 2 \int \! \int \!\{\widehat{\gamma}_{0}(s,t)- \gamma_{0}(s,t) \}^{2}\mathrm{d}s\mathrm{d}t\\
&= o_{\mathbb{P}}(1) + o_{\mathbb{P}}(1). 
\end{align*}
Therefore, $\|\widehat{\Gamma -\Gamma_{0}} - ( \Gamma- \Gamma_{0}) \|\overset{\mathbb{P}}{\to}0 $.
%$\widehat{\Gamma -\Gamma_{0}}= \widehat{\Gamma} -  \widehat{\Gamma}_{0}$ is a consistent estimator of $\Gamma- \Gamma_{0}$, 

We now turn to the $\Gamma_{0}^{-1} $ operator. We observe that $\Gamma_{0}^{-1}$ is an integral operator with kernel $\widehat{k}^{-1}(s,t)= \sum_{j=1}^{p} \widehat{\lambda}_{j}^{-1} \widehat{v}_{j} (s) \widehat{v}_{j} (t)$.  

For  $g,f\in L^{2} ([0,1]\times[0,1])$, we use the notation  $(gf)(s,t):= g(s,t) f(s,t)$. Then, we observe that
\begin{align*}
\int \! \int \!\{(\widehat{c}_{-0} \widehat{k}^{-1}) (s,t) &- (c_{-0} k^{-1})(s,t) \}^{2}\mathrm{d}s\mathrm{d}t\\
&=
\int \! \int \! \left[ \{ (\widehat{c}_{-0} -c_{0})  \widehat{k}^{-1}\} (s,t) +  \{c_{-0}(\widehat{k}^{-1} - k^{-1}) \} (s,t)\right]^{2}\mathrm{d}s\mathrm{d}t,
\end{align*}
with $k^{-1} (s,t)$ being the kernel with the true values $\lambda_{j}$ and $v_{j}$. Thus, 
\begin{align*}
\int \! \int \!\{(\widehat{c}_{-0} \widehat{k}^{-1}) (s,t) - (c_{-0} k^{-1})(s,t) \}^{2}\mathrm{d}s\mathrm{d}t \leq & 2 \int \! \int \! \{ (\widehat{c}_{-0} -c_{0})^{2}  (\widehat{k}^{-1})^{2}\} (s,t)\mathrm{d}s\mathrm{d}t  \\
+& 2 \int \! \int \! \{c_{-0}^{2}(\widehat{k}^{-1} - k^{-1})^{2} \} (s,t) \mathrm{d}s\mathrm{d}t.
\end{align*}
The first component on the right-hand side of the above inequality is bounded by $\int \! \int \! (\widehat{c}_{-0} -c_{0})^{2} (s,t) \mathrm{d}s\mathrm{d}t \int \! \int \!  (\widehat{k}^{-1})^{2} (s,t)\mathrm{d}s\mathrm{d}t $. If $\Gamma_{0}$ is invertible or if $\mathcal{H}$ is finite dimensional, then we have that $\widehat{\Gamma}^{-1}_{0}$ is bounded, and as a consequence $a_{1}:=\int \! \int \!  (\widehat{k}^{-1})^{2} (s,t)\mathrm{d}s\mathrm{d}t <\infty$. Then, 
\begin{align*}
2 \int \! \int \! \{ (\widehat{c}_{-0} -c_{0})^{2}  (\widehat{k}^{-1})^{2}\} (s,t)\mathrm{d}s\mathrm{d}t & \leq 2 a_{1} \int \! \int \! (\widehat{c}_{-0} -c_{0})^{2} (s,t) \mathrm{d}s\mathrm{d}t \\
& = 2 a_{1} o_{\mathbb{P}}(1).
\end{align*}
Similarly, for the second component we have that
\begin{align*}
 2 \int \! \int \! \{c_{-0}^{2}(\widehat{k}^{-1} - k^{-1})^{2} \} (s,t) \mathrm{d}s\mathrm{d}t & \leq 2 a_{2}  \int \! \int \! (\widehat{k}^{-1} - k^{-1})^{2} (s,t) \mathrm{d}s\mathrm{d}t \\
 & \leq 2 a_{1}^{2} a_{2}   \int \! \int \! (\widehat{k} - k )^{2} (s,t) \mathrm{d}s\mathrm{d}t \\
  & = O(1)o_{\mathbb{P}}(1),
\end{align*}
with $a_{2}:=  \int \! \int \! c_{-0}^{2}(s,t) \mathrm{d}s\mathrm{d}t  <\infty$. Therefore, we conclude that
$$
\int \! \int \!\{(\widehat{c}_{-0} \widehat{k}^{-1}) (s,t) - (c_{-0} k^{-1})(s,t) \}^{2}\mathrm{d}s\mathrm{d}t \overset{\mathbb{P}}{\to} 0, 
$$
that is, $\| \widehat{\Lambda} - \Lambda \| \overset{\mathbb{P}}{\to} 0$.

In the case where  $\mathcal{H}$ is infinite dimension, one can obtain similar result following similar ideas as in \cite[Ch.\ 8.3]{Bosq2000}. We omit the proof of this case.

 \hfill $\Box$
 \medskip

\noindent \textbf{Proof of Corollary \ref{C1}:}  
Let $\widehat{\Pi}_{F}$ and $\Pi_{F}$ be the orthonormal projectors on $\widehat{\mathcal{H}}_{F}$ and $\mathcal{H}_{F}$, respectively. Then, for any $z\in \mathcal{H}$ with $\|z\|_{\mathcal{H}}<\infty$, we have that 
\begin{align*}
\|\widehat{\Pi}_{F} (z) - \Pi_{F}(z) \| & \leq \sum_{k=1}^{K} \| \langle z, \widehat{F}_{k} \rangle  \widehat{F}_{k}  - \langle z, F_{k} \rangle  F_{k}   \| \\
& \leq a_{3} a_{4}  \sum_{k=1}^{K} \| \widehat{F}_{k} - F_{k} \| \\
& \leq a_{3} a_{4}  \sum_{k=1}^{K} \frac{2 \sqrt{2}}{\tilde{\alpha}_{k} } \| \widehat{\Lambda} - \Lambda \|,
\end{align*}
with $a_{3}= \max \{\langle z, \widehat{F}_{k} \rangle: \, k=1, \ldots, K \}$, 	$a_{4}= \max \{\langle z, F_{k} \rangle: \, k=1, \ldots, K \}$, $\tilde{\alpha}_{1}= \alpha_{1} -\alpha_{2}$, and $\tilde{\alpha}_{j}=\min\{ \alpha_{j-1} -\alpha_{j},   \alpha_{j} -\alpha_{j+1}\}$, $j=2, \ldots, K$. Since $\alpha_{1} > \alpha_{2}> \ldots > \alpha_{k} >0,$ we obtain $\|\widehat{\Pi}_{F} (z) - \Pi_{F}(z) \| \leq a_{3}a_{4} \frac{K 2 \sqrt{2}}{ \tilde{\alpha}} \| \widehat{\Lambda} - \Lambda\|$ with  $\tilde{\alpha}= \min \{ \tilde{\alpha}_{1}, \ldots,  \tilde{\alpha}_{K} \}$. Thus,  $\|\widehat{\Pi}_{F} (z) - \Pi_{F}(z) \| \leq O(1) o_{\mathbb{P}} (1)$, and then $\|\widehat{\Pi}_{F}  - \Pi_{F} \| \overset{\mathbb{P}}{\to} 0$. 

\hfill $\Box$
\medskip

\noindent \textbf{Proof of Proposition \ref{Prop:CS}:} We need only consider the case in which $r$ is $0<r <K$. We have that the long-run covariance operator $\Lambda_{\Delta X}$ of the functional process $\Delta X_{n} $ is $\Lambda_{\Delta X}=\Phi \Gamma_{\epsilon_{0}} \Phi$, where $\Phi= \sum_{j\geq 0} \Phi_{j}$, with $\Phi_{j}$ compact and self-adjoint operators. Thus, under Assumption \ref{NonS}, $X_{n}$ can be written as
$$X_{n}= X_{0} + \Phi \left(\sum_{i \leq n} \epsilon_{i} \right) + \nu_{n},$$
where $\nu_{n}$ is a stationary functional time series. Let $\xi_{k}$ be an eigenfunction of $\Lambda_{\Delta X}$; then, $\xi_{k}\notin \mathrm{ker}(\Phi )$. From this, we can show that $\mathcal{H}_{\lambda}^N = \{\mathrm{ker}(\Lambda_{\Delta X})\}^{\bot}$ and $\mathcal{H}_{\lambda}^S = \mathrm{ker}(\Lambda_{\Delta X})$ (see \cite{BeareEtal2017} for more details).
\hfill $\Box$
\medskip

\noindent \textbf{Proof of Corollary \ref{C2}:} The proof is similar in spirit to the proof of Corollary \ref{C1}.
\medskip

\noindent \textbf{Proof of Corollary \ref{C3}:} For $\xi_{k}\neq 0$, an eigenfunction of $\Lambda_{\Delta X}$, we have that $$\langle X_{n}, \xi_{k} \rangle= \langle X_{0}, \xi_{k} \rangle + \langle \Phi \left(\sum_{i \leq n} \epsilon_{i} \right) , \xi_{k} \rangle + \langle \nu_{n}, \xi_{k} \rangle= X_{0}^{k} + \sum_{i\leq n} \epsilon_{i}^{k} + \nu_{n}^{k} ,$$
where $X_{0}^{k}=  \langle X_{0}, \xi_{k} \rangle$, $\epsilon_{i}^{k}=\langle \epsilon_{i}, \Phi \xi_{k} \rangle $, and $\nu_{n}^{k}= \langle \nu_{n}, \xi_{k} \rangle $. That is, $\{\langle X_{n}, \xi_{k} \rangle \}$ has a random walk component $ \sum_{i\leq n} \epsilon_{i}^{k}$ since $\Phi(\xi_{k})\neq 0$. Thus, for each eigenfunction $\xi_{k}\neq 0$ of $\Lambda_{\Delta X}$, the process $\{ \langle X_{n}, \xi_{k}\rangle \}= \{\beta_{n,k} \}$ is an $I(1)$ process, and replacing $\xi_{k}$ with their corresponding estimator $\widehat{\xi}_{k}$, we obtain that $\{\widehat{\beta}_{n,k} \}$ is also an $I(1)$ process. In contrast, if $v\in  \mathrm{ker}( \Lambda_{\Delta X})$, then $\{ \langle X_{n}, v \rangle \}=  \langle X_{0}, v \rangle + \{ \langle \nu_{n}, v \rangle\}$ is stationary. This completes the proof.

\hfill $\Box$

\section*{Appendix B. Comparison of factors}
\begin{figure}[H]
\begin{center}
\includegraphics[scale=.38]{Figs/Lambda1Yield.pdf}\hspace{-.2cm}
\includegraphics[scale=.38]{Figs/Lambda2Yield.pdf}\hspace{-.2cm}
\includegraphics[scale=.38]{Figs/Lambda3Yield.pdf}
\includegraphics[scale=.38]{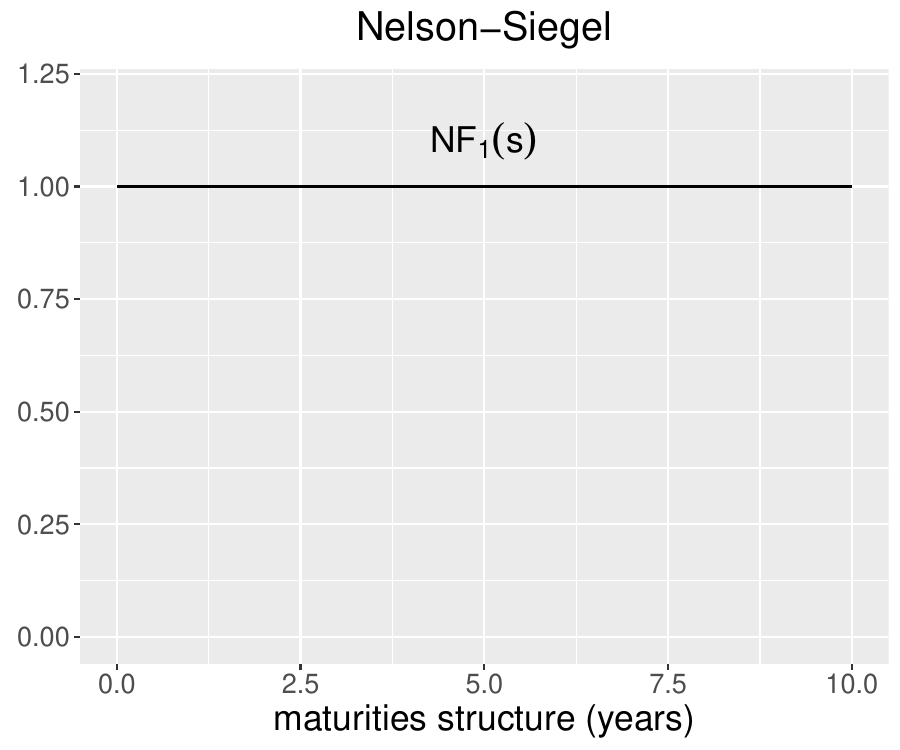}\hspace{-.2cm}
\includegraphics[scale=.38]{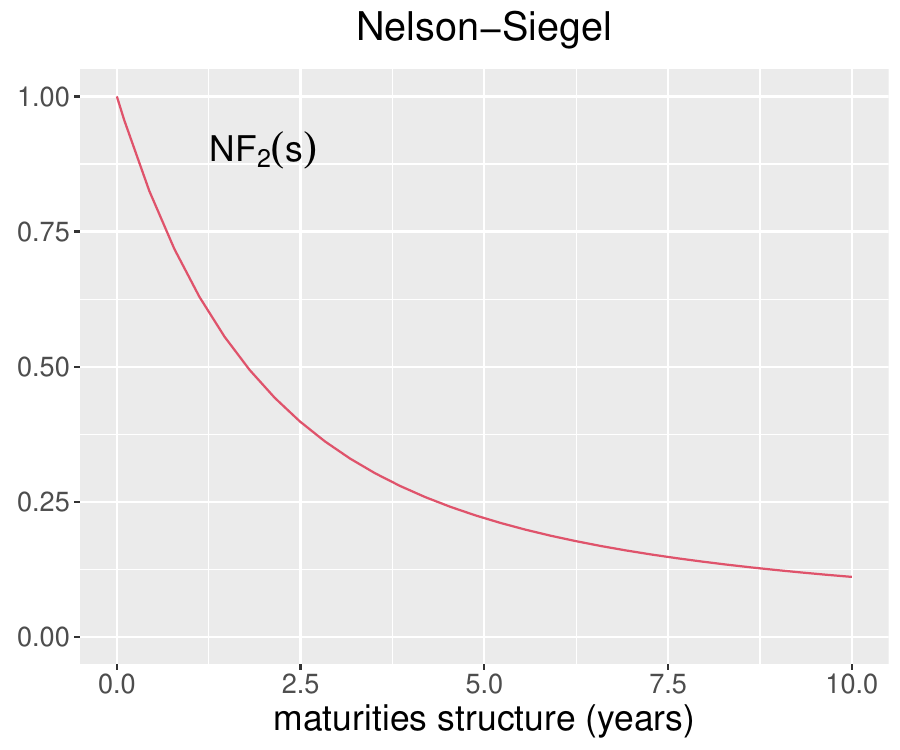}\hspace{-.2cm}
\includegraphics[scale=.38]{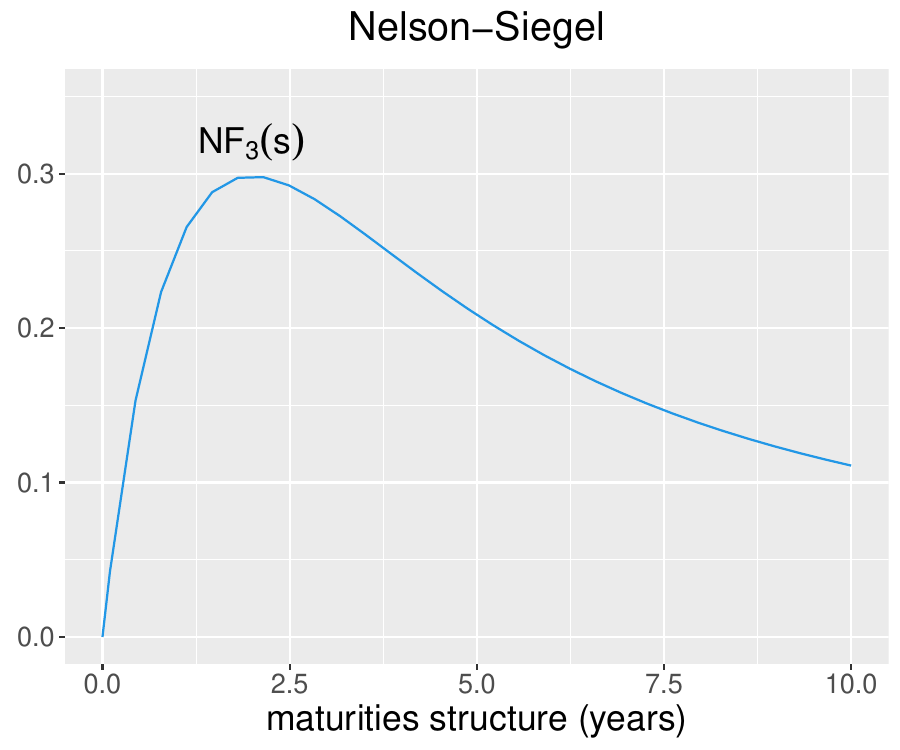}
\caption{Our estimator of factor loading curves, $\widehat{F}_{1}, \widehat{F}_{2},$ and $\widehat{F}_{3}$, and the Nelson-Siegel curves $NF_{1}$, $NF_{2}$, and $NF_{3}$. We fix $\lambda_{n}=0.9$.} \label{NScurves}
\end{center}
\end{figure}
Here we present the Nelson-Siegel curves corresponding to the three factors \cite[see, e.g.,][]{Diebold2006}. Also, we present our estimators of the Yield curve data (Section \ref{Application}). 

The  Nelson-Siegel curves are defined as 
$$
NF_{1} (s)=1, \,\, NF_{2}(s)= \frac{1- \exp(-\lambda_{n} s )}{\lambda_{n} s }, \,\, \mbox{and } NF_{3}(s)= \frac{1- \exp(-\lambda_{n} s )}{\lambda_{n} s } - \exp(-\lambda_{n} s).
$$
Figure \ref{NScurves} presents the three  Nelson-Siegel curves (second row) and our estimators (first row).  In the Nelson-Siegel curves, we fix the parameter $\lambda_{n}=0.9$.

\end{document}